\documentclass[aps,twocolumn,showpacs,superscriptaddress,unsortedaddress, showkeys]{revtex4-1}
\usepackage{graphicx}
\usepackage{amsmath}
\usepackage{amsfonts}
\usepackage{amsthm}
\usepackage{amssymb}
\usepackage{amsbsy}
\usepackage{pstricks}
\usepackage{wasysym}
\usepackage{mathrsfs}
\usepackage{subfigure}

\newcommand{\tq}{\tau_{\scriptscriptstyle Q}}

\newcommand{\Ai}{\operatorname{Ai}}

\newcommand{\vct}[1]{{\bf #1}}
\newcommand{\tk}{t_{\scriptscriptstyle Q}}
\newcommand{\lk}{l_{\scriptscriptstyle Q}}
\newcommand{\scale}[1]{\bar{#1}}

\newcommand{\sss}{\scriptscriptstyle}

\begin{document}

\title{The Kibble-Zurek Problem: Universality and the Scaling Limit}

\author {Anushya Chandran}
\affiliation{Department of Physics, Princeton University, Princeton, NJ 08544}
\author{Amir Erez}
\email[Corresponding author: ]{erezam@bgu.ac.il}
\affiliation{Department of Physics, Ben Gurion University of the Negev, Beer-Sheva 84105, Israel}
\author{Steven S.\ Gubser}
\author{S.\ L.\ Sondhi}
\affiliation{Department of Physics, Princeton University, Princeton, NJ 08544}

\date{\today}
\begin{abstract}
Near a critical point, the equilibrium relaxation time of a system diverges and any change of control/thermodynamic parameters leads to non-equilibrium behavior. The Kibble-Zurek problem is to determine the dynamical evolution of the system parametrically close to its critical point when the change is parametrically slow. The non-equilibrium behavior in this limit is controlled entirely by the critical point {\it and} the details of the trajectory of the system in parameter space (the protocol) close to the critical point.  Together, they define a universality class consisting of critical exponents---discussed in the seminal work by Kibble and Zurek---and scaling functions for physical quantities, which have not been discussed hitherto. In this article, we give an extended and pedagogical discussion of the universal content in the Kibble-Zurek problem. We formally define a scaling limit for physical quantities near classical and quantum transitions for different sets of protocols. We report computations of a few scaling functions in model Gaussian and large-$N$ problems and prove their universality with respect to protocol choice. We also introduce a new protocol in which the critical point is approached asymptotically at late times with the
system marginally out of equilibrium, wherein logarithmic violations to scaling and anomalous
dimensions occur even in the simple Gaussian problem.
\end{abstract}

\maketitle
\begin{picture}(0,0)(0,0)
\put(450,300){PUPT-2405}
\end{picture}

\section{Introduction}
\label{Sec:Introduction}

The study of critical points and their associated continuum limits or field theories has been a central and enormously productive exercise in statistical mechanics and condensed matter physics. This study began with the problem of equilibrium finite temperature transitions, was then extended to their dynamics and thereafter to the interplay between criticality and finite size effects \cite{Goldenfeld:1992aa}. This development has since been replayed in the theory of zero temperature quantum phase transitions \cite{Sondhi:1997aa, Sachdev:1999aa}. Needless to say, the development of powerful field theoretic methods, most notably conformal field theory in (1+1) dimensions \cite{Di-Francesco:1999aa} and most recently the gauge gravity duality \cite{Aharony:2000aa}, have significantly advanced this line of work. In all these cases the field theoretic approach not only describes the critical point but also the regions of the phase diagram adjacent to it.

A new dimension to the study of the passage through critical points was introduced by Kibble \cite{Kibble1976} in the context of the expanding universe, since recast in the language of critical phenomena by Zurek \cite{Zurek1985}. Their proposal, the ``Kibble-Zurek mechanism," is a theory of the defects generated in a system being cooled through a continuous symmetry-breaking phase transition at a small, but finite rate. The system inevitably goes out of equilibrium on the approach to the transition and arrives in the broken symmetry phase with different spatial regions realizing different orientations of the broken symmetry, and topological defects as a result. The mechanism predicts the scaling of the number of these defects with quench time \cite{Zurek1985, Zurek1996} and has been tested in a variety of systems \cite{Chen:2011aa, Hendry:1994aa, Ruutu:1996aa, Bauerle:1996aa, Maniv2003}, although quantitative agreement has been established only in zero dimensional annular Josephson junctions \cite{Monaco:2006ly} and in pattern-forming steady-state transitions \cite{Ducci:1999aa, Casado:2006aa} so there remains scope for more stringent experimental tests. It has been recently generalized to the setting of quantum phase transitions by Dziarmaga \cite{Dziarmaga:2005aa, Dziarmaga:2010aa}, Polkovnikov \cite{Polkovnikov:2005aa} and Zurek et al.\ \cite{Zurek:2005aa}, where the role of temperature is played by a non-symmetry breaking control parameter, and to ramps across multi-critical points \cite{Divakaran:2009fk, Divakaran:2008kx}. The scaling of other physical quantities like excess heat with quench time has also been investigated \cite{De-Grandi:2010aa}. Polkovnikov and coworkers have further paid attention to the interplay between finite system size and parameter velocity. 

The general problem posed by the work of Kibble and Zurek is that of a slow passage through a critical or multicritical point which we shall term the Kibble-Zurek (KZ) problem (often referred to by the oxymoronic term ``slow quench"). This problem is then characterized by a critical point with its equilibrium physics and a ``protocol," which is a particular path in the parameter space of the problem that touches the critical point. In the limit of asymptotically slow motion in parameter space, we expect that the physics is dominated by the critical point and hence is, in an appropriate sense, universal.  An important task of theory is then to isolate this universal content and compute it. We should note that the study of universality in the KZ problem, especially in the quantum setting, is a part of a wider current study of non-equilibrium quantum dynamics; for broader perspective see the recent discussion of Polkovnikov et al. \cite{Polkovnikov:2010aa}.

In previous studies of critical phenomena, two ideas have proven extremely useful. The first is the idea of the scaling limit in which various quantities of interest, such as thermodynamic densities and correlation functions, are postulated to obey certain homogeneity relations. This set of scaling functions along with the critical exponents captures the full universal content associated with a given critical point. The second is the renormalization group which provides an understanding of the origin of this universality and a full computation of its content.

In this paper we will make progress on the first front: we will formulate a scaling limit for the KZ problem and report model computations of the resulting scaling functions for a few classical and quantum problems. In appropriate limits, these scaling functions will reduce to those in the equilibrium and the coarsening problems. Except for recent work by Deng et al. \cite{Deng-S.:2008aa}, Biroli et al. \cite{Biroli:2010aa} and De-Grandi et al. \cite{De-Grandi:2011lq} discussing scaling functions in specific cases, the need for defining full scaling functions has been largely overlooked in the literature. Our contribution is to formalize the idea as a scaling limit for \emph{all} physical quantities for any pairing of a critical point and a protocol.

The second part of the program, beyond our ambition at present, would be the construction of a renormalization group flow. Such a program has been fruitfully pursued in sudden quench studies in classical models with stochastic dynamics (see \cite{Calabrese:2005ab} and references therein). We offer two modest steps in that direction here. First, in the classical context, we formulate the path-integral previously written down only for sudden quenches to the KZ problem. This is,
in principle, amenable to analysis by standard equilibrium renormalization group techniques.  Second, we will prove universality with respect to protocol choice for some model classical and quantum problems. Specifically, we will show that the expectation that only the behavior of the protocols in the vicinity of the critical points is important is, in fact, correct.

In the course of this paper we will offer a tripartite classification of possible protocols based on their ``topology"---i.e.\ whether they cross, turn around at or end at the critical point. The scaling of quantities such as the defect density and the excess heat with quench time has been previously generalized \cite{Sen:2008aa, De-Grandi:2010aa} to arbitrary \emph{positive} power-law behavior of the protocol on time near the critical point, that is, the first two kinds of protocols in our classification. Our contribution is the definition of a third class of power-law protocols that asymptotically end at the critical point and the identification of an interesting member---the marginal end critical protocol---that generates a one parameter family of non-equilibrium deformations of the equilibrium critical state worthy of further study.

We turn now to the organization of this paper. We begin, in Section~\ref{Sec:Classification}, by defining the
protocols of interest and introducing the well known KZ time and length.
In Section~\ref{Sec:ScalingLimitAllProtocols} we lay out the scaling formalism for the KZ problem---the definition of the KZ scaling limit and the resulting scaling functions. Here we also discuss how the KZ scaling
functions universally interpolate between the early equilibrium physics and the late time thermalization/coarsening physics. In Section~\ref{Sec:ClassicalModelA} we use the $O(N)$ vector model endowed with Model~A dynamics as our classical model system and compute scaling functions for it in the Gaussian and large-$N$ approximations. We also give an explicit proof of protocol universality for this problem. In Section~\ref{Sec:Quantum} we turn to the quantum $O(N)$ model in
the Gaussian approximation and compute various quantities of interest here. We conclude with
some remarks in Section VI.

We note that a recent paper by Kolodrubetz et al.\ reports analogous results for the transverse 
field Ising model in (1+1) dimensions \cite{Kolodrubetz:2011fj}. This work complements our own results on the quantum Ising universality class above the upper critical dimension ($3+1$) where the critical theory is Gaussian \footnote{Anticipating our classification, this statement is strictly true only for the
cis- and end-critical protocols. The dangerously irrelevant interaction needs to be included
to properly study the trans-critical protocols.}. We comment on the one case untreated in \cite{Kolodrubetz:2011fj} later in the paper. The present authors will also report computations of scaling forms in an interacting quantum field theory using the tools provided by the AdS/CFT correspondence elsewhere \cite{Chandran:2012aa}.

\section{Classification of Protocols and the KZ Time and Length Scales}
\label{Sec:Classification}

Consider a multicritical point in $d$ spatial dimensions.  Let $\{O_i\}$ denote the set of relevant operators that couple to conjugate fields $\{h_i\}$ and have scaling dimensions $\{\Delta_i\}$. At the critical point, we set $\langle O_i\rangle_{\rm eq}=0$. The set of scaling dimensions $\{\Delta_i\}$ and a host of scaling functions constitute the static universal equilibrium content of this critical point. Along with the dynamical exponent $z$ obtained from the time-dynamics, they determine the universality class of the critical point. Let $\delta$ parameterize a path in the space of conjugate fields $\{h_i\}$ such that $\delta$ is zero at the critical point. For every such path, define the correlation length exponent $\nu$ to be $1/(d-\Delta)$, where $\Delta$ is the scaling dimension of the most relevant operator that has a projection along the path near $\delta=0$. Along the path, the correlation length then diverges as $\delta^{-\nu}$ close to the critical point.

A useful example to keep in mind, particularly for the next section, is the ferromagnetic critical point. The relevant operators are the scalar energy operator and the vector magnetization, coupling to the conjugate fields of temperature and magnetic field respectively.

Unless mentioned otherwise, all length/time-scales are dimensionless and are measured in units of some microscopic length/time. We also set $\hbar$ and $k_B$ to $1$.

\subsection{Protocols in $\delta$}
\begin{figure*}[htbp]
\begin{center}
\includegraphics[width=5cm]{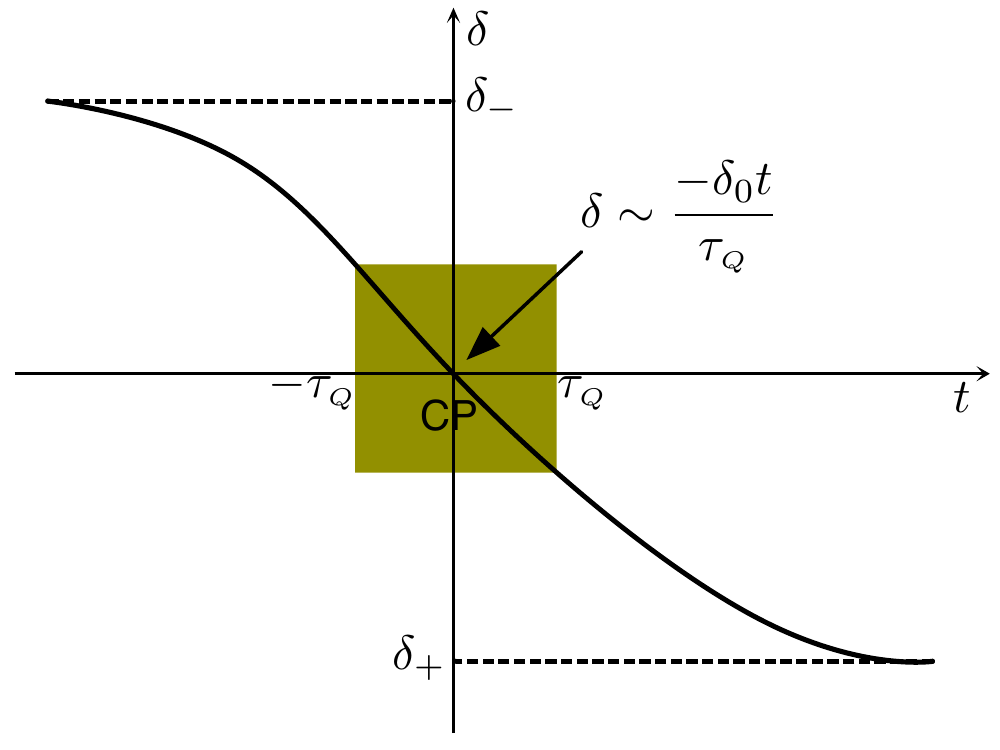}
\includegraphics[width=5cm]{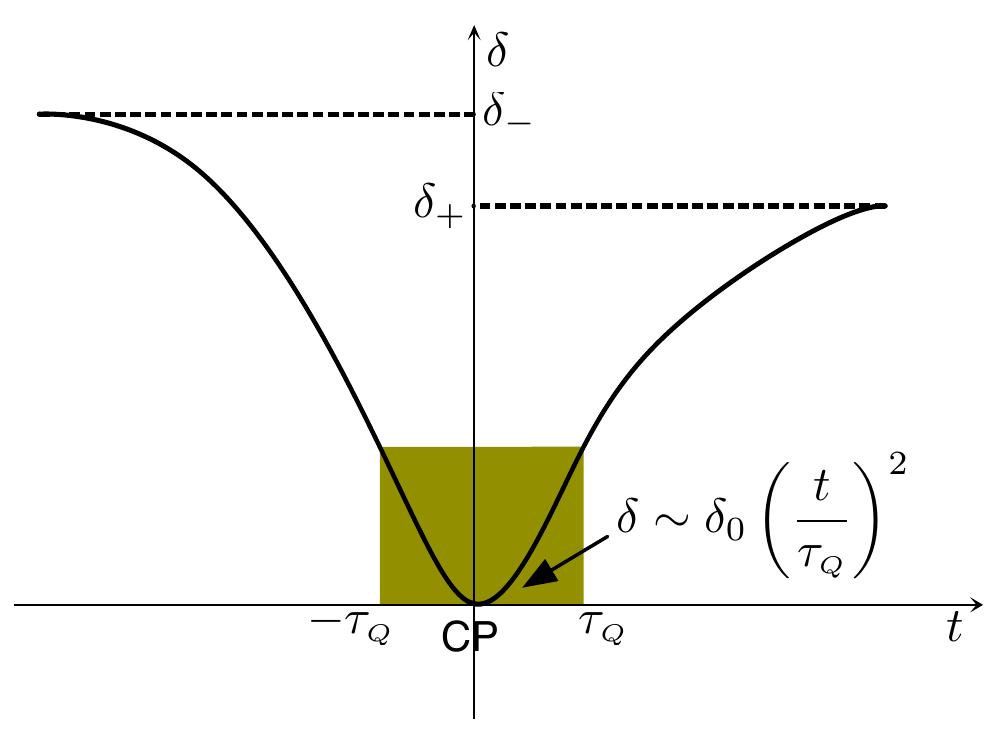}
\includegraphics[width=5cm]{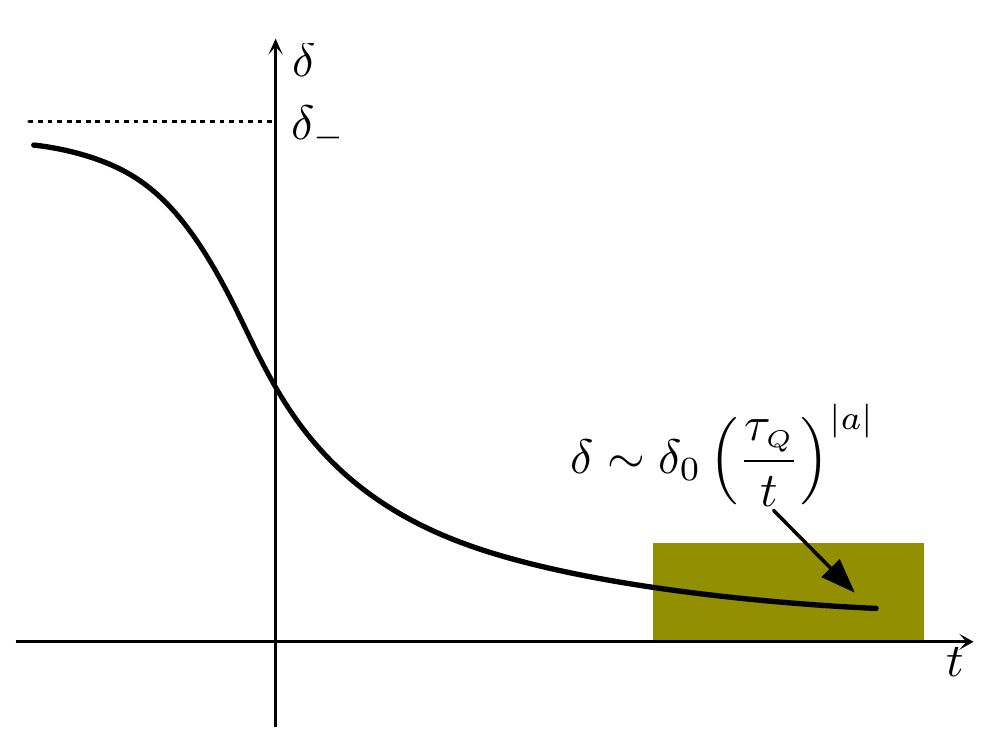}
\caption{From left to right, examples of Trans-Critical (TCP), Cis-Critical (CCP) and End-Critical (ECP) protocols. The critical point is at $t=0$. The leading order expansion of $\delta$ is valid in the shaded region. }
\label{Fig:protocols_cartoon}
\end{center}
\end{figure*}

Consider a system prepared in equilibrium at $t= -\infty$ at some fixed distance $\delta_->0$ away from the critical coupling $\delta=0$ being evolved in time along the path $\delta(t)$ in conjugate field space. For simplicity, we restrict ourselves to paths with a unique tangent most relevant operator near $\delta=0$. The KZ dynamics refines the classification of critical points discussed above using two pieces of data. The first is the symmetry of the path. If the most relevant operator along the path respects all the symmetries of the critical theory, the path is non-symmetry breaking, else it is symmetry-breaking. The second is the leading order behavior of $\delta(t)$ near the critical point  (strictly speaking, the critical coupling) which we classify below. 

\begin{itemize}
\item Trans-Critical Protocols (TCPs): These protocols take the system across the critical point. They smoothly interpolate $\delta$ between $\delta_->0$ as $t\rightarrow -\infty$ and $\delta_+<0$ as $t\rightarrow \infty$, crossing the critical point at $t=0$. An example of a TCP is
\begin{equation}
\label{Eq:translineartanheg}
\delta(t;\tq)=-\delta_0\tanh\frac{t}{\tq} \,.
\end{equation}
We will shortly show that the dynamic scaling functions are universal with respect to the behavior of the protocol near $t=0$. In anticipation of this result, we classify the entire family of analytic TCPs by their leading order behavior in a time-scale $\tq$ near $t=0$ as
\begin{align}
 \label{Eq:cistransprotocol}
 \delta(t;\tq)=\left\{\begin{array}{ll}
  \delta_-& t\rightarrow-\infty \\
  \delta_0\left(-\frac{t}{\tq}\right)^a \quad & t \rightarrow 0  \\
  \delta_+  & t\rightarrow\infty
 \end{array}
 \right.
 \end{align}
 with $a$ odd. $a=1$ is the linear protocol, $a=3$ the cubic and so on.
 \item Cis-Critical Protocols (CCPs): These protocols keep the system in a single phase and touch the critical point at $t=0$. They smoothly interpolate $\delta$ between $\delta_-$ as $t\rightarrow -\infty$ and $\delta_+>0$ as $t\rightarrow \infty$ through the critical point at $t=0$. An example of a CCP is
 \begin{align}
\delta(t;\tq)=\delta_0\tanh^2\frac{t}{\tq} \,.
\end{align}
Eq.~\eqref{Eq:cistransprotocol} with $a$ even classifies CCPs. $a=2$ is the quadratic protocol, $a=4$ the quartic and so on.

\item End-Critical Protocols (ECPs): 
End-Critical-Protocols (ECPs) keep the system in a single phase while asymptotically approaching the critical point. They smoothly interpolate between $\delta_-$ as $t\rightarrow -\infty$ and $0$ as $t\rightarrow \infty$. The asymptotic approach to the critical point may be with or without a time-scale. Here we restrict ourselves to the family of scale-free protocols:
\begin{align}
 \label{Eq:endprotocol}
 \delta(t)=\left\{\begin{array}{ll}
  \delta_-& t\rightarrow-\infty \\
   \delta_0 \left(\frac{t}{\tq}\right)^a=\delta_0 \left(\frac{\tq}{t}\right)^{|a|}\quad & t\gg \tq \,,
 \end{array}
 \right.
 \end{align}
where $a<0$ and $\tq$ is the time-scale over which the protocol behavior smoothly changes from being a constant to a power law.  Unlike in the above two cases, $a$ is not required to be integer valued.

\end{itemize}

\subsection{The KZ length and time}
\label{Sec:KZlengthtime}

A system evolving from $t=-\infty$ by a TCP, CCP or an ECP with large $|a|$ must fall out of equilibrium near the critical point due to critical slowing down. This is signaled by a diverging relaxation time $\xi_t$ \footnote{Near a quantum phase transition, this implies a vanishing many-body gap.}. The time at which the system falls out of equilibrium is defined to be the KZ time $\tk$. The KZ time defines a KZ length $\lk\sim \tk^{1/z}$. $\lk$ manifests in dynamic correlation functions as a crossover scale between equilibrium and non-equilibrium correlations. More intuitively \cite{Kibble1976, Zurek1996}, in an ordering transition, order is unable to form on scales larger than $\lk$ due to the finite quench rate, and domains of broken-symmetry phase of size $\lk$ persist in the ordered phase. We now separately consider the cases of TCP/CCP and ECP.

\emph{TCPs/CCPs.}  The scaling of the KZ length and time with $\tq$ follows essentially from considerations of \cite{Zurek:2005aa, Sen:2008aa}, which we recapitulate here.  A purely quantum version can also be formulated using Landau-Zener arguments and perturbation theory \cite{Zurek:2005aa, Polkovnikov:2005aa}. We define $\xi(t;\tq)$ and $\xi_t(t;\tq)$ to be the instantaneous correlation length and time if the system were in equilibrium at $\delta(t;\tq)$. The crucial quantities that determine $\tk$ are the change in the correlation time over a correlation time $\dot{\xi_t} \xi_t$ and $\xi_t$ itself. When $\dot{\xi_t} \xi_t \ll \xi_t$ or $\dot{\xi_t} \ll 1$, $\xi_t$ is changing slowly enough for the system evolution to be adiabatic. This is the case for $t<-\tq$. As $\dot{\xi_t}$ diverges at the critical point, there must come a time $-\tk>-\tq$ when it is of order one.\begin{align}
\dot{\xi_t}(\tk; \tq) &= 1 \nonumber \\
\Rightarrow \tk &\equiv \left(\frac{\tq}{\delta_0^{1/a}}\right)^{\frac{a\nu z}{a\nu z+1}} \,. \label{Eq:KZtime}
\end{align}

\emph{ECPs. } The asymptotic behavior of $\dot{\xi_t}$ categorizes the ECPs into three. $|a_c|\equiv1/(\nu z)$ below.
 \begin{itemize}
 \item Non-adiabatic : When $|a|>|a_c|$, $\dot{\xi_t}$ diverges at the critical point. The system falls out of equilibrium at $\tk$ given by Eq.~\eqref{Eq:KZtime} where $a$ is now negative.
 \item Adiabatic : When $|a|<|a_c|$, $\dot{\xi_t}$ is zero at the critical point. When $\tq$ is large, the evolution is adiabatic for all times.
  \item Marginal : When $|a|=|a_c|$, $\dot{\xi_t}$ is independent of $t$ at the critical point and the only length scale in the problem is $t^{1/z}$. The system is marginally out of equilibrium on this scale in a sense that will become clear soon.
  \end{itemize}
As we are interested in universal behavior \emph{out of equilibrium}, $|a|\geq |a_c|$ henceforth.

\section{The KZ Scaling limit}
\label{Sec:ScalingLimitAllProtocols}

\begin{figure}[htbp]
\begin{center}
\includegraphics[width=8cm]{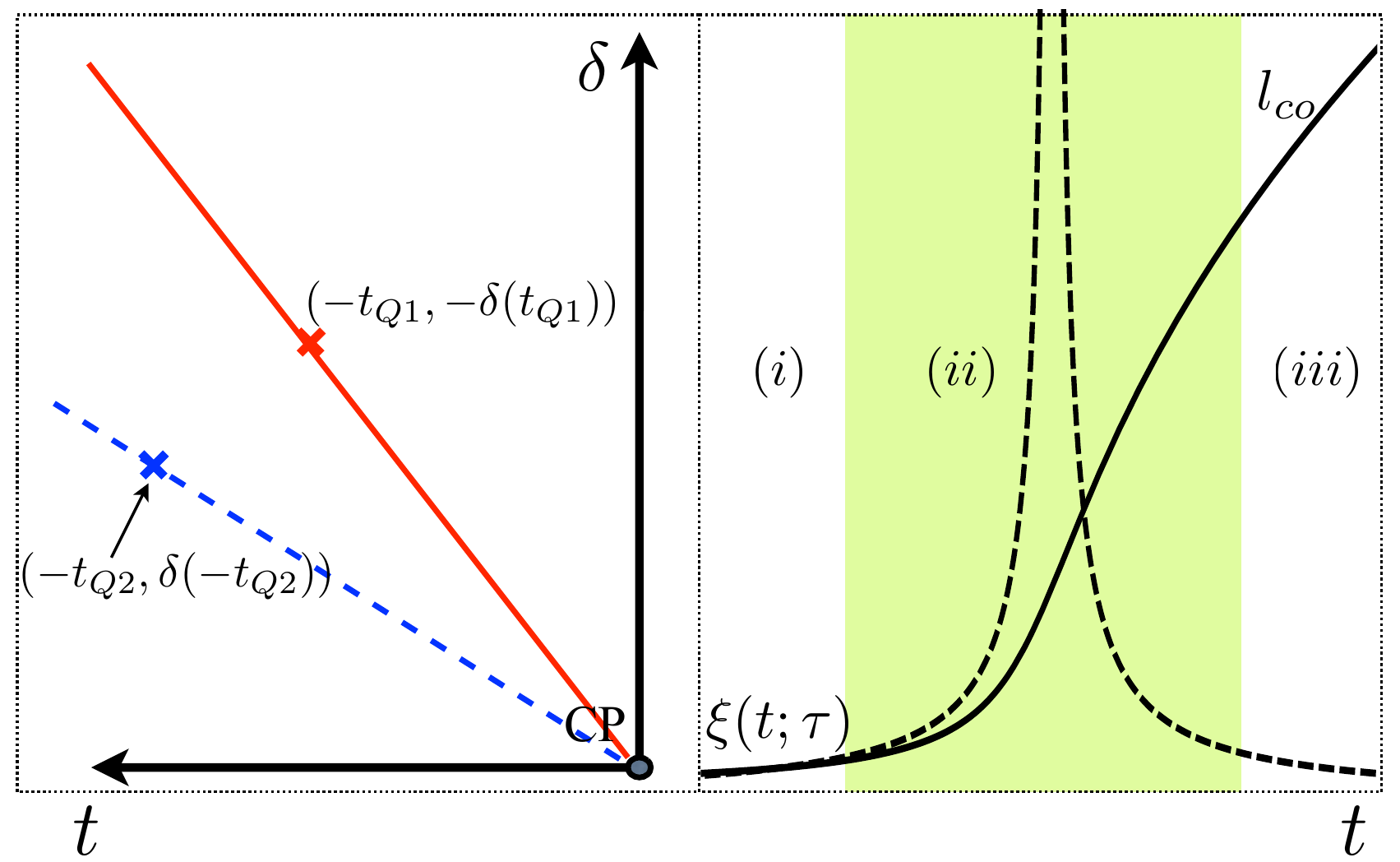}
\caption{Left: $\delta$ as a function of $t$ close to $t=0$ with KZ times shown for a slow (2,dashed) and fast (1,solid) quench. In the slower quench, the system falls out of equilibrium earlier ($|t_{\scriptscriptstyle Q2}|>|t_{\scriptscriptstyle Q1}|$) but at a smaller distance from the critical point ($\delta(-t_{\scriptscriptstyle Q2})<\delta(-t_{\scriptscriptstyle Q1})$). Right: The correlation length of a system quenched at a finite rate (solid) vs $\xi(t;\tq)$ (dashed). KZ split dynamics into (i) Adiabatic (ii) Sudden and (iii) Post-quench (here coarsening) regimes. The KZ scaling limit describes (ii) with (i) and (iii) as asymptotes. }
\label{Fig:2tau}
\end{center}
\end{figure}

The introduction of $\tk$ and $\lk$ is reminiscent of the introduction of finite size cutoffs in the theory of equilibrium critical behavior, and we are led to analogs of the finite size scaling limit and finite size scaling functions. We define the KZ scaling limit to be the limit $\tq\rightarrow\infty$ when time and length scales are measured in units of the diverging KZ scales, $\tk$ and $\lk$. $\delta(t;\tq) \rightarrow0$ in this limit, and the system is arbitrarily close to the critical point, evolving non-adiabatically for all $t/\tk$ (the shaded region in Fig.~\ref{Fig:2tau}b). The marginal ECP is special and we discuss it separately at the end of this section. 

We now turn to the definition of the scaling functions that arise in the KZ scaling limit. We discuss the scaling functions of  various physical quantities and their asymptotic forms in specific cases. Any readers new to scaling theory are encouraged to arm themselves with the Model A example in the Gaussian limit discussed in Sec.~\ref{Sec:ClassicalModelA} and interpret the definitions below in that context \cite{Goldenfeld:1992aa}. We initially consider classical critical points and discuss asymptotic forms when the transition is ordering. We then comment on the special features of quantum critical points.

Our notation is to denote the absolute value of any vector $\vct{k}$ by $k$, and to use $\scale{x}$ and $\scale{t}$ respectively to refer to the scaled length and time, $x/\lk$ and $t/\tk$. We reserve calligraphic lettering for scaling functions.  

\subsection{TCPs and CCPs}

\subsubsection{Scaling forms of correlation functions}
\label{Sec:scalingcorrfunc}

Consider the scalar operator $O$ with scaling dimension $\Delta$. We assume that the theory has translational and rotational invariance \footnote{With weak disorder, the KZ scaling forms apply to disorder averaged connected correlation functions. When the disorder distributions are broad, the typical moments satisfy the proposed scaling.}. The KZ scaling forms for the one and two-point connected correlation functions are
{\allowdisplaybreaks
\begin{align}
\label{Eq:2ptneqscalingform}
\langle O(\vct{x},t)\rangle_{\tq} &\equiv G_{\sss O} (t;\tq) \sim \frac{1}{\lk^\Delta} \mathcal{G}_{\sss O} \left(\frac{t}{\tk}\right) \nonumber \\
\langle O(\vct{x},t) O(\vct{x'},t')\rangle _{\tq} &\equiv  G_{\sss OO}(|\vct{x-x'}|,t,t'; \tq) \nonumber \\
&\sim
\frac{1}{\lk^{2\Delta}} \mathcal{G}_{{\scriptscriptstyle OO}}
\left(\frac{|\vct{x-x'}|}{\lk},\frac{t}{\tk},\frac{t'}{\tk}\right) \,.
\end{align}
When we use $\sim$ to indicate a scaling form of a correlation, as in (\ref{Eq:2ptneqscalingform}), we have a precise limiting statement in mind. For example, in the case of the two-point connected correlator, what we mean is
\begin{align}
\label{Eq:2ptneqscalingformexpl}
 \lim_{{\tq} \to^* \infty}
 \lk^{2\Delta} G_{\sss OO} (x,t,t';\tq)
=  \mathcal{G}_{{\scriptscriptstyle OO}}\left(\frac{x}{\lk},\frac{t}{\tk},\frac{t'}{\tk}\right) \,,
\end{align}
where $\tq \to^* \infty$ means the limit where $\tq \to \infty$ with $\frac{x}{\lk}$, $\frac{t}{\tk}$ and $\frac{t'}{\tk}$ held fixed. 

The scaling forms of all higher order cumulants and cross-correlators with other relevant operators are straightforward extensions of the form in Eq.~\eqref{Eq:2ptneqscalingform}. Note that $\mathcal{G}_{\sss O}$ can be identically zero. This is the case, for instance, in a zero-field temperature quench through the ferromagnetic critical point when $O$ is the spin operator. The finite ramp rate prevents order from forming on scales longer than the KZ length, and $\langle O\rangle_{\tq}$, the average magnetization, remains zero at all times. 

\subsubsection{Asymptotic form near equilibrium}
By construction of the protocol, we should recover equilibrium scaling forms in certain limits as $\scale{t}\rightarrow \pm \infty$. However, the precise limits are subtle and we derive them below. Recall that the equilibrium scaling limit is the limit of $\delta\rightarrow 0$ holding $x/\xi$ and $(t-t')/\xi^z$ fixed, wherein the two-point correlator has the scaling form,
\begin{align}
\label{Eq:GEqForm}
\langle O(\vct{x},t)  O(\vct{0},t') \rangle_\delta^{\rm eq}  &\sim \xi^{-2 \Delta}\mathcal{G}_{\sss OO}^{ \rm eq} \left(\frac{x}{\xi}, \frac{t-t'}{\xi^{ z}}\right) \,.
\end{align}
The $\sim$ symbol here is distinct from that in the previous subsection.

We consider the limit in which the KZ scaling form Eq.~\eqref{Eq:2ptneqscalingformexpl} reduces to the equilibrium scaling form Eq.~\eqref{Eq:GEqForm}. First, we observe that the relation,
\begin{align}
\label{Eq:xi_lk_t_tk}
\frac{x/\lk}{x/\xi(t;\tq)} = \frac{\xi(t;\tq)}{\lk}=\left|\frac{t}{\tk}\right|^{-a\nu},
\end{align}
implies that $x/\xi$ is fixed whenever $\scale{x}$ and $\scale{t}$ are held fixed. Similarly, $t/\xi^z$ and $t'/\xi^z$ are also fixed by $\scale{t}$ and $\scale{t'}$. Thus, an alternative to the KZ scaling form Eq.~\eqref{Eq:2ptneqscalingformexpl} involving the same two arguments as the equilibrium scaling form may be given:
\begin{align}
 \lim_{\tq \to^* \infty}
 \xi^{2\Delta} G_{\sss OO} (x,t,t';\tq)
=  \mathcal{G}_{{\scriptscriptstyle OO}}^{(2)}\left(\frac{x}{\xi},\frac{t-t'}{\xi^z},\frac{t+t'}{\tk}\right) \,.
\end{align}

When $|t| \gg \tk$, the system is in instantaneous equilibrium on length and time-scales, $\xi$ and $\xi_t$ respectively. This is the limit in which $\mathcal{G}_{\sss OO}^{(2)}$ must reduce to $\mathcal{G}_{\sss OO}^{\rm eq}$. Thus, in the limit $(\scale{t}+\scale{t'})\rightarrow \pm \infty$ holding $x/\xi$ and $(t-t')/\xi^z$ fixed:
\begin{align}
\mathcal{G}_{{\scriptscriptstyle OO}}^{(2)}\left(\frac{x}{\xi},\frac{t-t'}{\xi^z},\frac{t+t'}{\tk}\right) \sim \mathcal{G}_{{\scriptscriptstyle OO}}^{\rm eq}\left(\frac{x}{\xi},\frac{t-t'}{\xi^z} \right).
\end{align}
For the original KZ scaling form Eq.~\eqref{Eq:2ptneqscalingformexpl}, this translates to the requirement,
\begin{align}
 \label{Eq:FormalGEq}
\mathcal{G}_{\sss OO} (\scale{x},\scale{t},\scale{t'}) &\sim \scale{t}^{2 a \nu \Delta} \,\mathcal{G}_{\sss OO}^{\rm eq} (\scale{x} \,\scale{t}^{\nu a}, (\scale{t}-\scale{t'}) \,\scale{t}^{z a \nu}) \,.
\end{align}
in the same limit.
As expected,  $\xi(t;\tq)/\xi(t';\tq)\rightarrow 1$ so that there is a single diverging length in the equilibrium system. Furthermore, time-translation invariance is recovered.

In the example of a temperature quench through a ferromagnetic critical point, the equilibrium equal-time connected spin-spin correlation function decays exponentially on a length scale $\xi$ on either side of the critical point; consequently, Eq.~\eqref{Eq:FormalGEq} when $\scale{t}=\scale{t'}$ must asymptote to
\begin{equation*}
\mathcal{G}_{\sss OO} (\scale{x},\scale{t},\scale{t}) \sim \scale{t}^{2 a \nu \Delta} \,\exp(-\scale{x}\scale{t}^{\nu a}) \,.
\end{equation*}

\subsubsection{Asymptotic form for coarsening dynamics}
It is generally believed \footnote{For a recent dissent, see Olejarz et al.\ in Phys. Rev. E \textbf{83}, 05144 (2011) who have noted that the 3d Ising Model does not coarsen at zero temperature. More generally, the KZ scaling forms asymptote to the long-time behavior in the sudden quench.} that a system quenched to an ordered phase with multiple vacua undergoes coarsening , whereby each local broken-symmetry region grows in time and the system is asymptotically statistically self-similar on a characteristic length scale, $l_{\rm co}(t)\gg \xi$. Put another way, the two point function heals to its equilibrium value on the scale $\xi$ within each ``domain", and is exponentially suppressed between domains, each of growing length $l_{\rm co}\gg \xi$. In the late time regime, dynamical scaling is expected to hold when there are no growing scales competing with $l_{\rm co}$. For example, as $t,t'\rightarrow \infty$,
\begin{align}
\langle O(\vct{x},t)O(\vct{0},t') \rangle_\delta  &\sim \xi ^{-2 \Delta}\mathcal{G}_{\sss OO}^{\rm co} \left(\frac{x}{l_{\rm co}(t)}, \frac{x}{l_{\rm co}(t')}\right), \label{Eq:DynamicScaling} \\
\textrm{  where  } l_{\rm co}(t) \equiv t^\theta &=t^{ -a\nu+\frac{a\nu z+1}{z_d}} \nonumber.
\end{align}
$z_d$ is a dynamic exponent specific to coarsening. The coarsening scaling forms in various models are reviewed in \cite{Bray1994}.

When present in the KZ problem, we expect coarsening physics to emerge deep in the ordered phase, i.e. as $\scale{t},\scale{t'}\rightarrow \infty$, on the length scales $l_{\rm co}(t), l_{\rm co}(t')$. Proceeding as in the previous sub-section, we conclude that as $\scale{t},\scale{t'} \rightarrow \infty$ holding $x/l_{\rm co}(t)$ and $x/l_{\rm co}(t')$ fixed, the two-point function must have the limiting form:
\begin{align}
\label{Eq:GRealCoarsening}
&\mathcal{G}_{\sss OO} (\scale{x},\scale{t},\scale{t'}) \sim \scale{t}^{2 a \nu \Delta} \mathcal{G}_{\sss OO}^{\rm co} \left(\frac{\scale{x}} {\scale{t}^{\theta}}, \frac{\scale{x}} {\scale{t'}^{ \theta}}\right). \nonumber
\end{align}
Note that the limiting equilibrium form requires holding $x/\xi(t)$ fixed and that $l_{co}(t)/\xi(t)$ diverges as $\scale{t}\rightarrow\infty$.

In the simplest case of Model A dynamics for spins with $N$-components, $z_d=z=2$ and $l_{\rm co}(t) \sim \sqrt{t}$ in the infinite-$N$ limit. Although this phenomenology holds for systems with and without topological defects, the specific form of the equal-time spin-spin correlation function depends on the presence of topological defects \cite{Bray1994}.

\subsubsection{Scaling form of the non-equilibrium correlation length}
\label{Sec:CorrlengthDef}
The non-equilibrium correlation length $\xi_{\rm ne}$ is defined to be the inverse of the decay constant on the longest length scales of the two-point equal-time correlator in real space. In equilibrium, this length is the cross-over scale in correlation functions between fluctuations dominated by one fixed point and another. A particularly simple definition of $\xi_{\rm ne}$ is,
\begin{align}
\xi_{\rm ne}(t;\tq)=\sqrt{\frac{\int dx^d\, x^2 G_{\sss OO} (x,t,t;\tq)}{\int dx^d\, G_{\sss OO} (x,t,t;\tq)}},
\end{align}
and is useful when $G_{\sss OO}$ is always positive.
A more general definition is through the smallest imaginary part of the poles of $G_{\sss OO}(k,t,t;\tq)$ in $k$-space; we will elaborate on this when the necessity arises. Biroli and coauthors \cite{Biroli:2010aa} recently discussed the KZ scaling form of $\xi_{\rm ne}$ in non-symmetry breaking TCPs. They observed that $\xi_{\rm ne}$ must asymptote to the equilibrium correlation length, $\xi(t;\tq)$, as $\scale{t}\rightarrow -\infty$; that it must scale according to the critical coarsening form, $\tk^{1/z}$, when $|\scale{t}|\sim O(1)$; and that it must asymptote to the coarsening length $l_{\rm co}$ as $\scale{t}\rightarrow\infty$. Their proposed scaling form can be derived from that of $G_{\sss OO}$ and can be re-written as
\begin{align}
\xi_{\rm ne}(t;\tq) \sim \lk \mathcal{L}_{\rm ne}\left(\frac{t}{\tk}\right).
\end{align}
The asymptotic behavior of $\xi_{\rm ne}$ for $\hat{t} \to \pm\infty$ formally translates to the limits
\begin{align}
 \label{Eq:xineasymptotes}
 \mathcal{L}_{\rm ne}(x)\sim\left\{\begin{array}{ll}
  |x|^{-a\nu}\quad& x\rightarrow-\infty \\
  |x|^\theta & x\rightarrow\infty.
 \end{array}
 \right.
 \end{align}
Absent coarsening physics, the right asymptote reproduces the instantaneous correlation length.

\subsubsection{Scaling form of the number of defects}

An intuitively appealing picture of the lack of order on length scales greater than the KZ length
is through topologically protected point defects of characteristic separation $\lk$, and/or defects of dimension $p$ with characteristic separation $\lk$ in any hyperplane of co-dimension $p$. This picture is really only meaningful if the separation between defects is much larger than the equilibrium correlation length; a constraint met only in the coarsening regime, $t \gg \tk$. In the coarsening regime, the density of a defect of dimension $p$ in the hyperplane of co-dimension $p$ should scale as $1/l_{\rm co}(t;\tq)^{d-p}$. At fixed positive $\scale{t}$, this reproduces the celebrated scaling of the defect density with $\tq$,
\begin{align}
\textrm{Density of defects } \sim \left(\frac{1}{\lk}\right)^{d-p} \sim\tq^{\frac{a\nu(d-p)}{a\nu z+1}}.
\end{align}
The above scaling with $\tq$ has been verified in experiments in $0$-dimensional annular Josephson junctions \cite{Monaco:2006ly} and in non-linear optical and hydrodynamical systems undergoing steady-state transitions \cite{Ducci:1999aa, Casado:2006aa}. It is worth noting that the frequently cited experiments \cite{Chuang:1991aa, Bowick:1994aa} in liquid crystal systems do not test this critical scaling, but instead test the coarsening scaling forms. Although we have presented the above as a natural scaling ansatz, it may be derivable from the known scaling of correlation functions, for example, by the methods of Halperin-Liu-Mazenko for classical transitions \cite{Halperin:1981aa, Liu:1992fk}.

\subsubsection{Scaling forms of thermodynamic analogs}
\label{Sec:ClassicalAction}
The dynamics of classical systems is typically modeled phenomenologically by stochastic differential equations, possibly with conservation laws \cite{Hohenberg:1977aa}. It is often useful to reformulate the stochastic dynamics in $d$-dimensions in terms of a path integral in $(d+1)$ \cite{Hochberg1999,Martin1973}. For Gaussian noise, the generating functional of correlation functions of the fundamental field $O(x,t)$ is
\begin{equation*}
Z[J]  = \int \mathcal{D}O(x,t) \exp\left(-\int d^dx dt\, (\mathcal{L} + J(x,t) O(x,t) )\right).
\end{equation*}
Here $J$ is the source for $O$, and $\mathcal{L}$ is a local Lagrangian density in which parameters of the protocol (like $\tq$) appear as couplings. When hyper-scaling is obeyed, all KZ scaling forms follow from the scaling of the associated free energy.

In this formulation, it is natural to define a time-dependent free energy density. We divide the $(d+1)$-dimensional space-time into a stack of spatial slices with volume $L^d$ and temporal length $\Delta t$.  Neglecting boundary effects, we  compute the free-energy density of a slice:
\begin{align}
f(t;\tq) =\lim_{L\rightarrow \infty \atop \Delta t\rightarrow 0} \frac{-\log(Z)}{L^d \Delta t}
\end{align}
where
\begin{align*}
Z(t) =   \int \mathcal{D}O \exp\left(-\int_{\sss \vct{x}\in[-L,L]^d} d^dx \int_{t}^{t+\Delta t} dt \,\mathcal{L}\right) \,.
\end{align*}
For time-independent $\delta$, $f$ is also time-independent, and its leading non-analytic dependence on $\delta$ is $f_{\rm na}\sim \delta^{\nu d}$. When $\delta$ varies with time, we conjecture a scaling form for $f_{\rm na}(t;\tq)$ in the KZ limit :
\begin{equation}
\label{Eq:fscalingform}
f_{\rm na}(t;\tq) \sim \frac{1}{\lk^d} \mathcal{F}\left(\frac{t}{\tk}\right).
\end{equation}
When $t/\tk\ll-1$, the system is asymptotically in equilibrium. Thus,
\begin{equation}
\mathcal{F}\left(\frac{t}{\tk}\right) \sim \left(\frac{t}{\tk}\right)^{a\nu d}  \textrm{  when  } t/\tk \ll -1.
\end{equation}
This is also the expected asymptotic form of $\mathcal{F}$ if the system does not coarsen for $t/\tk\gg 1$.

\subsection{ECPs}
\label{Sec:MarginalECPScalingTheory}
The definition of the scaling limit, the KZ scaling forms of various observables and their asymptotic behavior when the system is subjected to a non-adiabatic ECP closely follows the TCP discussion. Qualitatively:
\begin{itemize}
\item When $\scale{t}\rightarrow 0$, the system evolution is adiabatic and equilibrium scaling forms are recovered on the scale of the instantaneous correlation length and time. 
\item The long-time behavior of a sudden quench to the critical point is recovered in the limit $\scale{t}\rightarrow\infty$. On a growing length scale $l_{\rm co}(t)=t^{1/z}$, the system appears self-similar and satisfies dynamical scaling (Eq.~\eqref{Eq:DynamicScaling}). On length scales $x$ much smaller than $l_{\rm co}(t)$, correlations have relaxed to their critical form, while for $x \gg l_{\rm co}(t)$ the system coarsens. 
\item Scaled times near $1$ probe a universal early-time regime of a sudden quench, i.e. $t \ll \xi_{t,0}$ where $\xi_{t,0}$ is the relaxation time before the quench. In the sudden quench, this regime is tied to boundary criticality \cite{Janssen:1989kx}.   
\end{itemize}

When $|a|\nu z=1$ in Eq.~\eqref{Eq:endprotocol}, $\dot{\xi_t}$ is a constant for $t>\tq$. By suitably re-defining constants, Eq.~\eqref{Eq:endprotocol} can be re-written as :
\begin{align}
 \label{Eq:mecp}
 \delta(t)=\left\{\begin{array}{ll}
  \delta_-& t\rightarrow-\infty \\
   \left(\frac{\theta}{ t}\right)^{1/\nu z}\quad & t\gg \tq \,.
 \end{array}
 \right.
 \end{align}
The key distinction between the marginal ECP and all the other protocols is that there is only one growing length scale $t^{1/z}$ everywhere in the power-law regime. At late times, we expect the system to appear self-similar on this scale. This leads us to the scaling limit :  $x, t\rightarrow \infty$ at fixed $\theta$ holding $x/t^{1/z}$ fixed.
\begin{align}
\label{Eq:2ptneqscalingformecp}
G_{\sss O} (t) &\sim \frac{1}{t^{\Delta/z}} \nonumber\\
G_{\sss OO}(x,t,t') &\sim
\frac{1}{t^{2\Delta/z}} \mathcal{G}_{{\scriptscriptstyle OO}}
\left(\frac{x}{t^{1/z}},\frac{t}{t'}\right).
\end{align}
We emphasize that the scaling functions above do not connect to the adiabatic limit. They are distinct from $\mathcal{G}_{\sss OO}^{eq}$. Unfortunately, this also implies that they necessarily contain some non-universal data. The source of this non-universality will be clearer in Sec.~\ref{Sec:GaussianECP}, where we are nevertheless able to identify interesting non-equilibrium behavior. 

\subsection{Quantum systems}
We comment below on the major differences that arise in the treatment of quantum systems.
\begin{itemize}
  \item The scaling of the correlation functions proceeds as before. However, the analogs of thermodynamic quantities of interest for protocols that begin with the system at $T=0$ are now the excitation energy density in excess of the energy density in the adiabatic ground state, or ``heat'' density \cite{De-Grandi:2010aa} \footnote{See \cite{Polkovnikov:2008ab} for a discussion of why the total excitation energy is sensibly called heat. For infinite systems where one needs to work with intensive quantities though, a heat density is not a useful concept especially when macroscopic subregions exhibit thermal equilibration}, and the entropy density for which plausible definitions can be constructed from the diagonal entropy \cite{Anatoli:2011aa, Polkovnikov:2008ab} or the entanglement entropy of a macroscopic subregion. These are respectively expected to exhibit the scaling forms
      \begin{align}
      \label{Eq:esscalingform}
      q(t;\tq) &\sim \frac{1}{\lk^{d+z}} \mathcal{Q}\left(\frac{t}{\tk}\right) \\
      s(t;\tq) &\sim \frac{1}{\lk^d} \mathcal{S}\left(\frac{t}{\tk}\right).
      \end{align}
      For an isolated quantum system, by construction of the protocol, $s$ and $q$ are constant as $t\rightarrow -\infty$. As $\scale{t}\to \infty$, $s$ and $q$ also tend to a constant provided the system thermalizes to the Gibbs or the generalized Gibbs ensemble. 
      
       \item For protocols that begin with the system in equilibrium at $T > 0$, a new dimensionless parameter, $k_B T/(\hbar/\tk)$,  now enters the quantum problem. Along with the quantities held fixed as the quench time is taken to be arbitrarily large in Sec.~\ref{Sec:scalingcorrfunc}, we also hold $T \tk$ fixed. For example, the scaling form for the 2-point unequal time correlation function is now
\begin{align*}
\langle O(0,t) O(\vct{x'},t')\rangle _{\tq,T} \sim \frac{1}{\lk^{2\Delta}} \mathcal{G}_{\sss OO}\left(\frac{x'}{\lk},\frac{t}{\tk},\frac{t'}{\tk},T\tk\right).
\end{align*}
The definition of the entropy goes through as before but the excitation energy density is now measured with respect to the energy density that would be obtained in a strictly adiabatic evolution \footnote{For an isolated system at finite temperature, the dynamics involves starting with a typical state with the thermodynamic limit energy density. At high temperatures, a typical state must look ``classical''; this suggests that the behavior of the entanglement entropy and issues of many-body localization \cite{Basko:2006aa} need further examination.}.

%

   \item Integrable quantum systems allow the existence of sharply defined quasiparticle excitations; their density resolved by momentum can serve as a uniquely quantum observable for CCPs and ECPs. When the vacuum is not unique, it is not straightforward to define quasiparticles and separate them from domain walls and other topological defects. Currently, we are not aware of an integrable system above one dimension where this question can be properly posed \cite{Uhlmann:2010aa} (in one dimension there is no real distinction between quasiparticles and domain walls). Note that the definition of the thermodynamic quantities of heat and entropy do not rely on integrability. For instance, in a cyclic process like a CCP, $q(t;\tq)$ when $t\gg \tk$ is the difference of the system's energy density between symmetric time points.
\end{itemize}

\subsection{Comments}

We conclude this section with four comments.

\begin{itemize}
  \item The Kibble-Zurek picture was initially proposed for a linear ramp through a thermal transition. The time-dynamics was decoupled into three regions: the adiabatic regime for $t<-\tk$, the diabatic or sudden or impulse region from $-\tk$ to $\tk$ when the system is frozen, and the post-quench regime for $t>\tk$ when the system is unfrozen and evolves through domain growth \cite{Zurek1996}, defect-anti-defect annihilation etc. Recently \cite{Biroli:2010aa}, this picture was extended to account for evolution in the impulse regime using critical coarsening results. The KZ scaling forms introduced in this article also probe dynamics in the impulse regime with the adiabatic and the post-quench regimes acting as asymptotic limits.

 \item A finite system dimension $L$ can be readily accommodated in our scaling forms in the combination $L/\lk$. The results obtained through adiabatic perturbation theory \cite{Polkovnikov:2005aa, Polkovnikov:2008aa} can therefore be fitted within this framework.

\item We expect that the KZ scaling forms are unchanged by the inclusion of interaction terms in the dynamics that are irrelevant to the critical theory \footnote{We exclude dangerously irrelevant operators from this discussion as they will modify the asymptotic behavior in the KZ scaling limit.}. In the classical setting, the reformulation of the dynamical problem in $d$-dimensions as an inhomogenous statistical problem in $(d+1)$-dimensions allows us to establish this by standard power-counting arguments.  The quantum case is more subtle and needs to be treated case-by-case.

\item As the free Gaussian theory is the stable critical point in $d>4$, the KZ scaling forms that we calculate below in that theory are universal for $d>4$. In the classical setting, interaction terms in the Gaussian theory can be treated as perturbations to the action defined in Sec.~\ref{Sec:ClassicalAction} and will modify dynamics only on time-scales $t_{int}$ that are parametrically larger than $\tk$. In the quantum case, these terms induce scattering between the free quasi-particles of the unperturbed theory on time-scales  $t_{int}$ that are parametrically larger than $\tk$. In either case, $t_{int}/\tk\rightarrow\infty$ in the scaling limit and the effects of these terms drop out.

\end{itemize}

\section{Classical Systems with Model A dynamics}
\label{Sec:ClassicalModelA}
We illustrate the universality of the KZ scaling limit and explicitly compute scaling functions of a vector operator $\vec{ \phi}$ in the simplest setting: Model A \cite{Hohenberg:1977aa} dynamics with a Landau-Ginzburg-Wilson (LGW) free energy. Model A dynamics is dissipative and obeys no conservation laws. Let $\vec{\phi}$ be an $N$-component vector field in $d=3$ dimensions. The dimensionless LGW free-energy and the equation of motion are, respectively,
\begin{align}
\label{Eq:ModelA}
F=\int d^3x \left[\frac{1}{2}\left(|\nabla \vec{\phi}|^2  +  r_0|\vec{\phi}|^2\right.\right. &+\left.  \left.\frac{u}{2N}|\vec{\phi}|^4\right) - \sqrt{N} h^\alpha \phi_\alpha\right] \nonumber \\
\frac{\partial \phi_\alpha}{\partial t} = - \frac{\partial F}{\partial \phi_\alpha} + \zeta_\alpha \,.
\end{align}
$x$ and $t$ are dimensionless and measured in units of the inverse cutoff $\Lambda^{-1}$ and $\Lambda^{-z}$ respectively. $\zeta_\alpha$ is a zero-mean spatially uncorrelated white-noise stochastic variable for every $\alpha$. The variance of $\zeta_\alpha$ is chosen to be $2$ so that the long-time limit of the structure factor computed from the equation of motion when $F$ is time-independent is equal to the equilibrium structure factor:
\begin{align}
\label{Eq:Noisestats}
\langle \zeta_\alpha(\vct{x},t) \zeta_\beta(\vct{x'},t')\rangle = 2 \delta_{\alpha\beta} \delta^3(\vct{x}-\vct{x'}) \delta(t-t') \,.
\end{align}

\subsection{Gaussian Limit}
In this limit, we drop the $\vct{\phi}^4$ term in $F$. The critical point is at $r_0=0,h^\alpha=0$, i.e.\ at the origin in the $(r_0,h^\alpha)$ parameter space. Let us restrict ourselves to paths in this space that lie along the axes and include the origin. The equilibrium theory is sensible only at the origin and when $r_0>0$. We can therefore study CCPs and ECPs along the non-negative $r_0$ axis ($\delta=r_0$) . Luckily, we can also study TCPs along the same axis as the time-dependent fields are finite even when $r_0<0$. This physically uninteresting protocol is pedagogically useful. The critical exponents for all such paths are $\nu=1/2$, $z=2$. In the remainder of the discussion, $\delta=r_0$ and $h^\alpha,u=0$.

Eq.~\eqref{Eq:ModelA} is linear in $\vec{\phi}$ and hence diagonal in Fourier space. It can be explicitly solved for any protocol $\delta(t;\tq)=r_0(t;\tq)$ and a fixed noise realization:
\begin{align}
\label{Eq:phiEOM}
\phi(\vct{k},t) = \int_{-\infty}^t dt' \, e^{-\int_{t'}^t dt''(k^2+r_0(t'';\tq))}\, \zeta(\vct{k},t') \,.
\end{align}
We have dropped the component label of $\vec{\phi}$ for brevity. Memory of the initial condition at $t=-\infty$ is lost on the time-scale $1/r_0(-\infty;\tq)$ and is therefore absent in the solution above. The equal time structure factor for each component, defined as $\langle \phi(\vct{k},t)\phi(\vct{k'},t)\rangle = (2\pi)^3 \delta^3(\vct{k}+\vct{k'}) G_{\sss \phi\phi}(k,t;\tq)$, is
\begin{align}
\label{Eq:GaussianG}
G_{\sss \phi\phi}(k,t;\tq) &= 2\int_{-\infty}^t dt' \, e^{-2\int_{t'}^t dt''(k^2+r_0(t'';\tq))} \,.
\end{align}

\subsubsection{Scaling limit for $\tanh$ TCP}
The universality of the scaling limit with respect to details of the the protocol is already apparent when we consider the simple TCP $r_0(t;\tq)=-\tanh(t/\tq)$. This is an example of a protocol that is linear near the critical point, i.e.\ $a=1$ in Eq.~\eqref{Eq:cistransprotocol}. Consequently, the KZ time and length scales are
\begin{align*}
\tk=  \sqrt{\tq} ,\,\, \lk= \tq^{1/4}.
\end{align*}
Re-writing Eq.~\eqref{Eq:GaussianG} in units $\scale{k}=k\lk$ and $\scale{t}=t/\tk$,
\begin{align*}
G_{\sss \phi\phi}(k,t;\tq) &= 2\tk\int_{-\infty}^{\scale{t}} d\scale{t}' e^{-2\int_{\scale{t}'}^{\scale{t}} d\scale{t}''(\scale{k}^2-\tk\tanh(\scale{t}''/\tk))} \,.
\end{align*}
As $\tq \rightarrow \infty$, $\tk\tanh(\scale{t}''/\tk) \rightarrow \scale{t}''$ and the KZ scaling form of the two-point equal time correlation function is
\begin{align}
G_{\sss \phi\phi}(k,t;\tq) &\sim \lk^2 \mathcal{G}_{\sss \phi\phi}(\scale{k}, \scale{t}) \nonumber \\
&\sim  \lk^2 \int_{-\infty}^{\scale{t}} 2d\scale{t}'  e^{-2\int_{\scale{t}'}^{\scale{t}} d\scale{t}''(\scale{k}^2-\scale{t}'')} \,. \label{Eq:GGaussianscaling}
\end{align}
 Observe that the scaling function only depends on the leading behavior of the protocol near the critical point. For all protocols such that $r_0(t;\tq)\sim -t/\tq$ in the vicinity of  zero time, the scaling form of the structure factor is the expression that we have just derived.
\subsubsection{Scaling functions for all TCPs and CCPs}
Wick's theorem informs us that all higher order cumulants of $\phi$ only depend on $G_{\sss \phi\phi}$. The details of the protocol enter the expression of $G_{\sss \phi\phi}$ only in the combination $\tk r_0(t;\tq)$. In the scaling limit, for any $a$ in Eq.~\eqref{Eq:cistransprotocol}, it is easily seen that
\begin{align}
\label{Eq:Smalltbehavior}
\tk r_0(t;\tq) \sim \left(\frac{-t}{\tk}\right)^a \,.
\end{align}
The scaling function of the equal time correlation function for any protocol with the near-zero behavior in Eq.~\eqref{Eq:cistransprotocol} is therefore
\begin{align*}
\mathcal{G}_{\sss \phi\phi}(\scale{k},\scale{t}) = 2\int_{-\infty}^{\scale{t}} d\scale{t}' \, e^{-2\int_{\scale{t}'}^{\scale{t}} d\scale{t}''(\scale{k}^2+(-\scale{t}'')^a)} \,.
\end{align*}
\begin{figure}[htbp]
\begin{center}
\includegraphics[width=8cm]{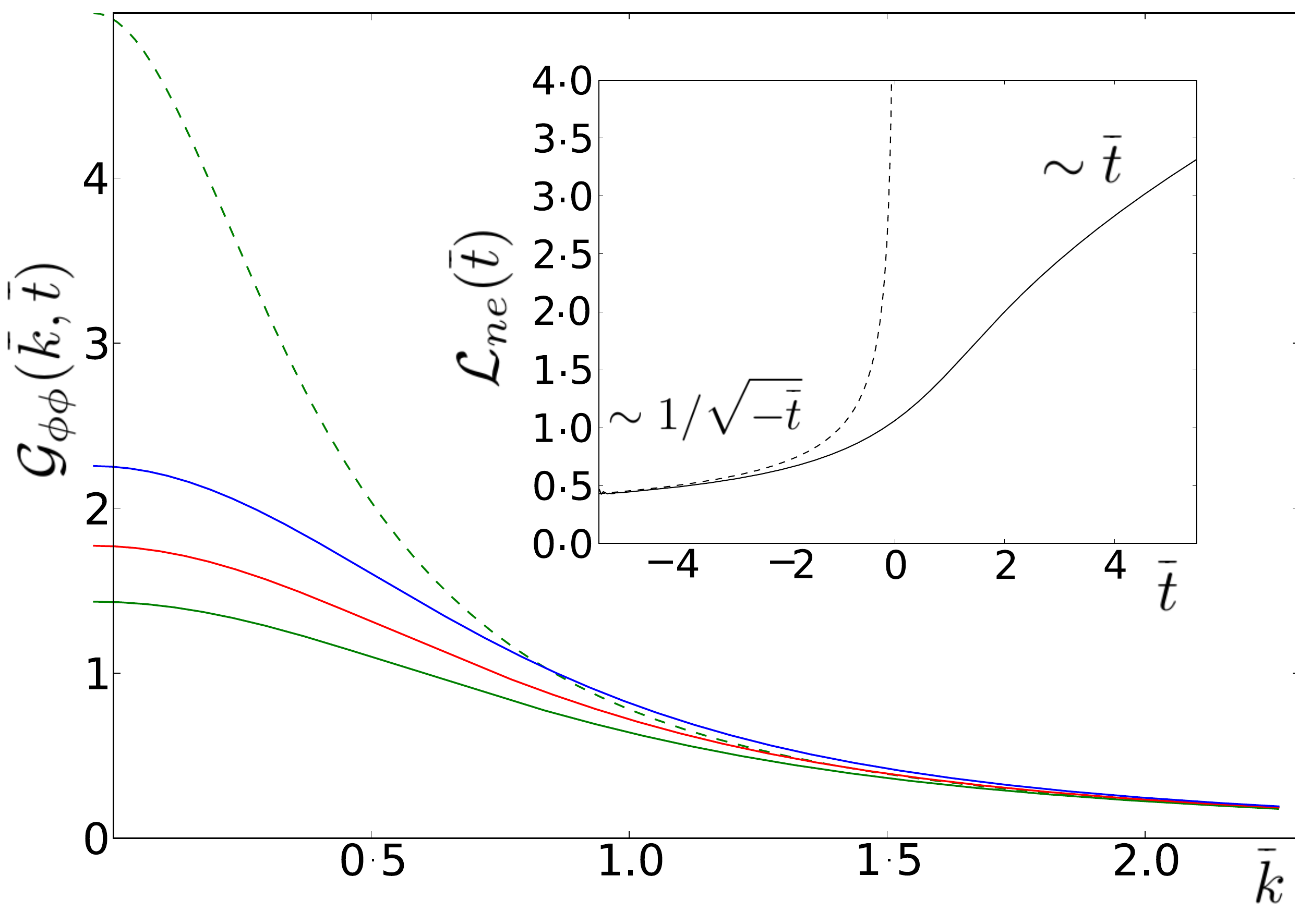}
\caption{$\mathcal{G}_{\sss \phi \phi}(\scale{k},\scale{t})$ vs $\scale{k}$ at fixed time slices for the linear TCP. The blue, red and green solid lines are respectively at $\scale{t}=0.2,0$ and $-0.2$. The green dashed line is the correlator if the system were in equilibrium at $\scale{t}=-0.2$. Inset: The scaling form of the non-equilibrium correlation length vs $\scale{t}$ in solid. The dashed line is the instantaneous correlation length in units of $\lk$, $\xi(t;\tq)/\lk$.}
\label{Fig:linear_gaussian}
\end{center}
\end{figure}

As seen from the Fig.~\ref{Fig:linear_gaussian} for the linear TCP, $\mathcal{G}_{\sss \phi\phi}$ and $\mathcal{L}_{\rm ne}$ match equilibrium forms as $\scale{t}\rightarrow -\infty$ and $\scale{k}\gg 1$:
\begin{align}
\lk^2 \mathcal{G}_{\sss \phi\phi} (\scale{k},\scale{t}) &\sim \xi^2 \,\mathcal{G}^{\rm eq}_{\sss \phi\phi}(k\xi(t))\equiv \frac{\xi(t)^2}{k^2\xi(t)^2 + 1} \nonumber \\
\xi_{\rm ne} &\sim \lk\mathcal{L}_{\rm ne} \equiv \frac{1}{(-t)^{1/2}} \,. \label{Eq:GEqasymptote}
\end{align}
Observe that $\mathcal{G}_{\sss \phi\phi}(0,0)$ is finite, indicating the suppression of order on length scales longer than $\lk$. As $\scale{t}\rightarrow \infty$, the non-equilibrium correlation length grows without bound for TCPs due to the inverted potential. For CCPs, on the other hand, all scaling forms asymptote to the equilibrium forms at large positive times.

In Appendix \ref{App:dplusonef}, we compute the partition function in (3+1) dimensions and demonstrate the validity of the scaling hypothesis for the free-energy density $f_{\rm na}$.
\subsubsection{Scaling functions for the ECPs}
The scaling functions of the non-adiabatic ECPs is identical to that of the TCPs/CCPs when $a<-1/\nu z$. Their asymptotic behavior may also be verified to be in agreement with Sec.~\ref{Sec:MarginalECPScalingTheory}. The more interesting case is the marginal ECP $r_0(t)=\theta/t$. Independent of the early-time regularization on the time-scale $\tq$, $G_{\sss \phi\phi}$ has the asymptotic form in Eq.~\eqref{Eq:2ptneqscalingformecp}:
\begin{align*}
G_{\sss \phi\phi}(k,t) &\sim t \,\mathcal{G}_{\sss \phi\phi}(k\sqrt{t}) \\
\mathcal{G}_{\sss \phi\phi}(x) &\equiv \frac{4 e^{-2x^2}}{x^{4\theta+2}} \int_0^{x} dy \, e^{2y^2} y^{4\theta+1}.
\end{align*}
As promised, $\mathcal{G}_{\sss \phi\phi}(x)$ is distinct from $\mathcal{G}_{\sss \phi\phi}^{\rm eq}(x)$ and the non-equilibrium correlation length is a multiple of the instantaneous one $\xi(t)=\sqrt{t/\theta}$ at late times. The system exhibits critical coarsening and relaxes to the critical point as $t\to\infty$.


\subsection{Large-$N$ limit}
In the infinite $N$ limit, the LGW theory is exactly solvable in terms of 1 and 2-point correlators. All higher order correlators follow by the application of Wick's theorem. The 1 and 2 point correlators are known exactly in equilibrium and can be reduced to quadrature with Model A dynamics. The theory in $d=3$ that we discuss here exhibits non-trivial critical behavior and affords us a probe of symmetry-breaking paths and coarsening physics.

Assume in Eq.~\eqref{Eq:ModelA} that $h^i\neq 0$ for $i=1$ and zero otherwise. This specifies all one-point correlators for $i>1$ to be zero. The two-point correlator for every component is the same and is a function of the variable $m^2=r_0 + u\langle \phi_\alpha^2\rangle$. The well-known self-consistency equations relating $\langle \phi_1 \rangle$, $G_{\sss \phi\phi}$ and $m^2$ are
\begin{align}
r_0 + u \left(\int^\Lambda \frac{d^3k}{(2\pi)^3} G_{\sss \phi\phi}(k) + \frac{\langle \phi_1 \rangle^2}{N} \right) = m^2 \label{Eq:SelfConsistencymass} \\
\langle \phi_1 \rangle = \frac{\sqrt{N} h^1}{m^2} \label{Eq:SelfConsistencyField} \,,
\end{align}
where $\Lambda$ is a cutoff on the maximum allowed $|k|$.

The third equation that completes the theory in equilibrium is,
\begin{equation}
\label{Eq:GLargeNEq}
G^{\rm eq}_{\sss \phi\phi}(k) = \frac{1}{k^2 + m^2} \,.
\end{equation}
The critical point in the equilibrium theory is at $r_0=r_c=-u\Lambda/2\pi^2$ and $h^\alpha=0$. The two relevant operators that couple to $r_0$ and $h^\alpha$ are respectively $|\vec{\phi}|^2$ and $\phi_\alpha$, with scaling dimensions $1$ and $5/2$ in $d=3$.  They are respectively non-symmetry breaking and symmetry-breaking. Consequently, we can study all three kinds of protocols along the $r_0$ axis ($\nu=1,z=2$) or along any of the $h^\alpha$ axes ($\nu=2/5,z=2$).

The equation of motion \eqref{Eq:ModelA} for Model A dynamics is
\begin{align}
\label{Eq:EOMLargeN}
\frac{\partial\phi_\alpha(\vct{k},t)}{\partial t} = -(k^2+m^2)\phi_\alpha(\vct{k},t) + h^\alpha (2\pi)^3 \delta^3(\vct{k}) + \zeta_\alpha \,.
\end{align}
This specifies the functional dependence of $G_{\sss \phi\phi}(k,t;\tq)$ on $m^2$ and completes the dynamical theory.

\subsubsection{Scaling limit for TCPs and CCPs along the $r_0$ axis}
\label{Sec:ScalingLargeNNSBTCP}
Here $h^\alpha=0$ and $\delta(t;\tq)=r_0(t;\tq)-r_c$. Eq.~\eqref{Eq:SelfConsistencyField} implies that the one-point correlator for all components is zero: $\langle \phi_\alpha \rangle =0$.  We henceforth drop component subscripts.  We will make frequent use of the Gaussian result \eqref{Eq:GaussianG}, which we reproduce here for the reader's convenience:
\begin{align} \label{Eq:Gform}
G_{\sss \phi\phi}(k,t;\tq;r_0) &=   2\int_{-\infty}^t dt' \, e^{-2\int_{t'}^t dt''(k^2+r_0(t'';\tq))} \, \nonumber
\end{align}
The notation $G_{\sss \phi\phi}(k,t;\tq;r_0)$ emphasizes that $G_{\sss \phi\phi}$ depends on $r_0(t;\tq)$.

Observe that the solution to Eq.~\eqref{Eq:EOMLargeN} is exactly that of the free Gaussian case---Eq.~\eqref{Eq:phiEOM}---with $r_0(t;\tq)$ replaced by $m^2(t;\tq)$. Consequently, the third equation that completes the dynamical theory is
\begin{align}
 G_{\sss \phi\phi}(k,t;\tq) = G_{\sss \phi\phi}(k,t;\tq;m^2) \,,
\end{align}
where the right hand side refers to the Green's function of~\eqref{Eq:Gform} with $r_0(t;\tq)$ replaced by $m^2(t;\tq)$.  The critical value $r_c$ of $r_0$ in this notation is,
\begin{align}
r_c = -u \int^\Lambda \frac{d^3 k}{(2\pi)^3} G_{\sss \phi\phi}(k,t;\tq;0) \,.
\end{align}
One can now re-express \eqref{Eq:SelfConsistencymass} as
\begin{align}\label{Eq:MSquaredConsistent}
m^2(t;\tq) &= \delta(t;\tq) + u \int^\Lambda \frac{d^3 k}{(2\pi)^3}
  \Big[ G_{\sss \phi\phi}(k,t;\tq;m^2)   \nonumber \\
  &{} - G_{\sss \phi\phi}(k,t;\tq;0) \Big] \,.
\end{align}

Let us now take the scaling limit. The KZ length and time scales are $\tk = (\tq^a/\delta_0)^{\frac{2}{2a+1}}$ and $\lk= \sqrt{\tk}$. Dimensional considerations imply that the scaling form of $m^2(t;\tq)$ is $m^2(t;\tq)\sim \mathcal{M}^2(\scale{t}) /\lk^2$. Using the leading order behavior of the protocol near the critical point, $\delta(t;\tq) \sim (-\scale{t})^{a}/\lk$, we derive the following scaled form of Eq.~\eqref{Eq:MSquaredConsistent}:
\begin{align}
\label{Eq:Scaledmeq}
 (-\scale{t})^a + 2u \int_{-\infty}^{\scale{t}} d\scale{t}' \int^{\infty} \frac{d^3 \scale{k}}{(2\pi)^3} &\left(e^{-2\int_{\scale{t}'}^{\scale{t}} d\scale{t}''(\scale{k}^2+\mathcal{M}^2)} \right.\nonumber \\
 &\left.-e^{-2\int_{\scale{t}'}^{\scale{t}} d\scale{t}''\scale{k}^2}\right)=0
 \end{align}
We note the following points:
 \begin{enumerate}
  \item The cutoff $\Lambda$ appears in the unscaled equation only as the upper limit on the momentum integral. In the scaling limit, the upper limit $\sim \Lambda \lk \rightarrow \infty$ and the scaled relation is \emph{cutoff-independent}.
  \item The scaled relation, being a limiting form, is \emph{simpler} than the unscaled one.
  \item The relation above is unchanged by keeping higher order terms in $r_0(t;\tq)$. This is tantamount to proving the universality of all correlators with respect to details of the protocol away from the critical point in the scaling limit.
 \end{enumerate}

\subsubsection{Scaling forms for the linear TCP}
\begin{figure}[htbp]
\begin{center}
\includegraphics[width=8cm]{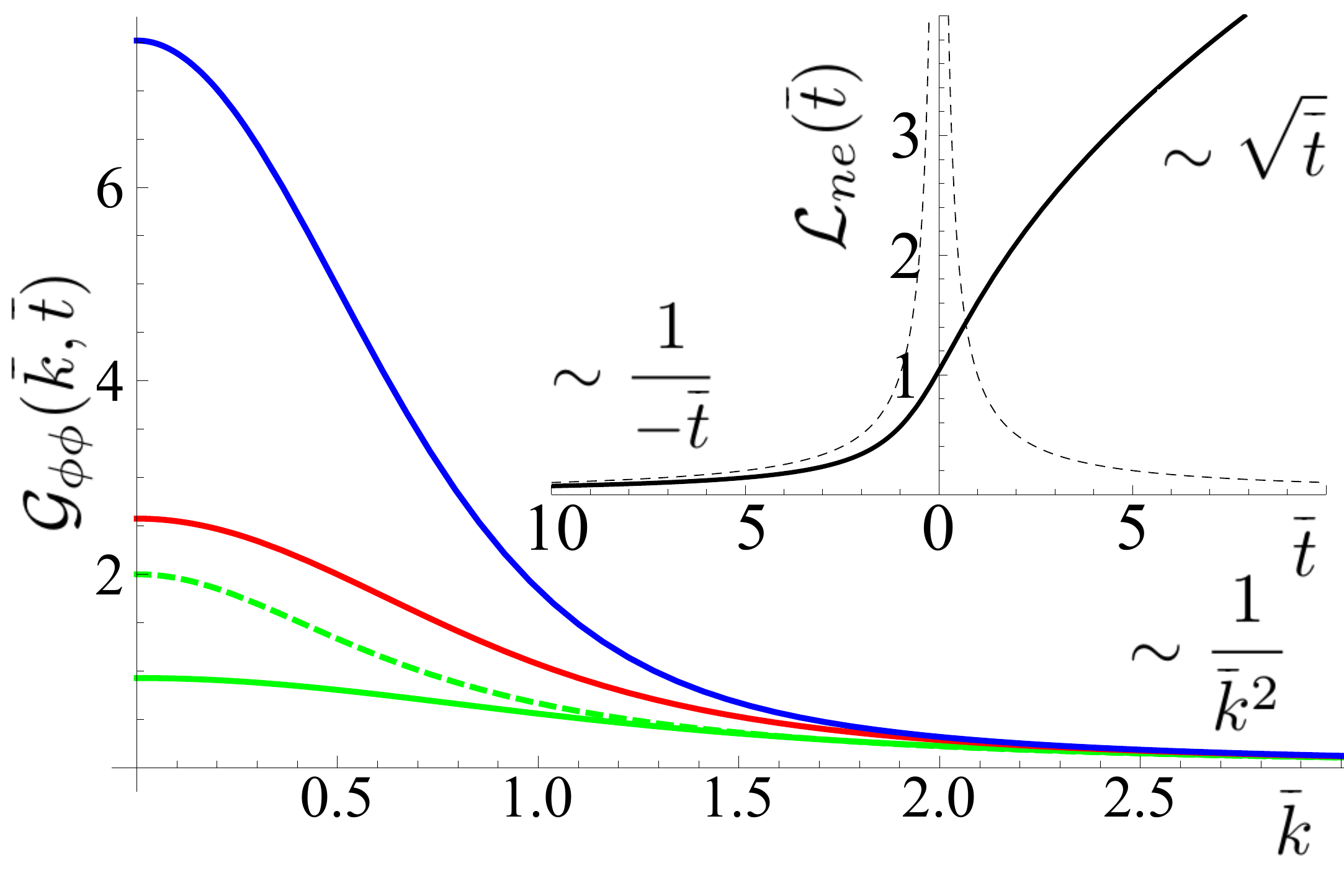}
\caption{$\mathcal{G}_{\sss \phi \phi}(\scale{k},\scale{t})$ vs $\scale{k}$ at fixed time slices for the linear TCP. The blue, red and green solid lines are respectively at $\scale{t}=0.5,0$ and $-0.5$. The green dashed line is the correlator if the system were in equilibrium at $\scale{t}=-0.5$. Inset: $\mathcal{L}_{\rm ne}$ vs $\scale{t}$ (solid) and $\xi/\lk$ (dashed).}
\label{Fig:linear_largen}
\end{center}
\end{figure}

Fortunately, Eq.~\eqref{Eq:Scaledmeq} can be solved when $a=1$. The choice $u=\sqrt{8}\pi$ simplifies pre-factors and makes apparent the form of the solution. Defining
\begin{align}
\label{Eq:Deff}
f(\scale{t}) \equiv e^{2\int_0^{\scale{t}} d\scale{t'}\mathcal{M}^2(\scale{t'})} \hspace{5pt}\textrm{or}\hspace{5pt}\mathcal{M}^2(\scale{t}) &= \frac{f'(\scale{t})}{2f(\scale{t})},
\end{align}
the solution is
\begin{align*}
f(\scale{t}) &= - 3^{1/3} \Gamma(1/3) e^{2\scale{t}^3/3} (\scale{t} \Ai(\scale{t}^2) + \Ai'(\scale{t}^2))\,.
\end{align*}
$\Ai$ is the Airy function of the first kind and $\Gamma$ is the gamma function. We may reduce $\mathcal{G}_{\sss \phi\phi}$ to quadrature:
\begin{align*}
\mathcal{G}_{\sss \phi\phi}(\scale{k},\scale{t}) = 2\int_{-\infty}^{\scale{t}} d\scale{t}'  e^{-2\int_{\scale{t}'}^{\scale{t}} d\scale{t}''(\scale{k}^2+\mathcal{M}^2(\scale{t}''))},
\end{align*}
and compute all higher cumulants using Wick's theorem.

As $\scale{t}\rightarrow-\infty$, we recover the equilibrium behavior in Eq.~\eqref{Eq:GEqasymptote} with $\xi\sim 1/(-t)$. The results from the coarsening literature \cite{Bray1994} in this theory are :
\begin{align*}
l_{\rm co}(t)&= \sqrt{t} \\
\mathcal{G}_{\sss \phi\phi}^{\rm co} (k l_{\rm co}(t) ) &= \exp(-2 (kl_{\rm co}(t))^2).
\end{align*}
We should therefore expect that $\mathcal{L}_{\rm ne} \sim \sqrt{\scale{t}}$ and $\mathcal{G}_{\sss \phi\phi} \sim \scale{t}^{5/2} \exp(-2\scale{k}^2 \scale{t})$ as $\scale{t}\rightarrow\infty$. As is seen from Fig.~\ref{Fig:linear_largen}, both asymptotes are correctly predicted for $\mathcal{L}_{\rm ne}$. We have checked this for $\mathcal{G}_{\sss \phi\phi}$ as well. This verifies that the asymptotes predicted in Eq.~\eqref{Eq:xineasymptotes} and in Eq.~\eqref{Eq:GRealCoarsening} are correct in this theory.

\subsubsection{Scaling limit for TCPs and CCPs along $h_1$ axis}
Setting $r_0=r_c$, we can explore symmetry-breaking protocols along the $h^1$ axis. The implicit relation from the self-consistency equations for the linear TCP is
\begin{align}
\left(\int\limits_{-\infty}^{\scale{t}} d\scale{t'} \, \scale{t'}\sqrt{f(\scale{t'})}\right)^2 = \int\limits_{-\infty}^{\scale{t}} d\scale{t'}\, \frac{f(\scale{t})-f(\scale{t'})}{|\scale{t}-\scale{t'}|^{3/2}}.
\end{align}
Although scaling is guaranteed, we cannot proceed further as the solution to this equation is not known.


\section{Gaussian Quantum Field Theories}
\label{Sec:Quantum}

We now turn to the KZ problem near quantum critical points. Again, it will prove instructive to investigate the class of ferromagnetic critical points with $O(N)$
symmetry. We will consider the simplest non-trivial case, that of Gaussian
scalar fields.  The Lagrangian we consider is,
\begin{equation}\label{Eq:QuantumLagrangian}
{\cal L} = {1 \over 2} \left[ \partial_\mu \phi \partial^\mu \phi - m^2 \phi^2 \right]  \,,
\end{equation}
where we permit $m^2$ to depend on time ($\delta\equiv m^2$) .  Standard critical exponents include $z=1$ (owing to relativistic symmetry when $m$ is a constant) and $\nu=1/2$ (owing to the Yukawa form of the Green's function for static sources when $m$ is a constant).

\subsection{Second quantization}

The second quantized treatment of $\phi$ is based on the expansion
\begin{equation}
\phi({\bf x},t) = \int {d^d k \over (2\pi)^d} e^{i {\bf k} \cdot {\bf x}} \phi_{\bf k}(t) \,,
\end{equation}
where we set
\begin{equation}
\label{Eq:phikdefinition}
\phi_{\bf k}(t) = f_{\bf k}(t) a_{\bf k} + f^*_{-{\bf k}}(t) a^\dagger_{-{\bf k}}
\end{equation}
and impose the commutation relations
\begin{equation}
[a_{\bf k}, a^\dagger_{\bf k'}] = (2\pi)^d \delta^d({\bf k} - {\bf k'}) \,.
\end{equation}
Owing to rotational symmetry, the mode functions $f_{\bf k}(t)$ and Green's functions only depend on the magnitude of the momentum $k$.
The mode functions $f_k(t)$ satisfy the mode equation
\begin{equation}\label{Eq:ModeEquation}
\left[ {d^2 \over dt^2} + \Omega_k^2(t) \right] f_k(t) = 0 \,,
\end{equation}
where
\begin{equation}
\Omega_k^2(t) \equiv k^2 + m^2(t) \,.
\end{equation}
One must also impose the Wronskian condition
\begin{equation}\label{eq:Wron}
f_k \dot f^*_k - \dot f_k f^*_k = i
\end{equation}
in order to obtain standard commutation relations between $\phi_{\bf k}$ and its conjugate $\pi_{\bf k} = \dot\phi^\dagger_{\bf k}$. Once the mode functions $f_k$ are specified, a Fock space vacuum $|0\rangle$ can be defined through the conditions $a_{\bf k}|0\rangle = 0$.

In general, Eq.~\eqref{Eq:ModeEquation} is hard to solve exactly. However, the techniques of WKB provide an approximate solution when $\Omega_k$ is real, positive and varying ``slowly enough"; to this end, let us define the oscillatory, positive frequency WKB solution:
\begin{equation}
\tilde{f}_k(t) = \frac{1}{\sqrt{2\Omega_k(t)}} \exp\left\{-i\int^t dt' \, \Omega_k(t') \right\} \,,
\label{eq:WKB_formal}
\end{equation}
where the lower limit of integration can be specified at our later convenience.  If one defines $\tilde{a}_{\bf k}(t)$ through the equations $\phi_k(t) = \tilde{f}_k(t) \tilde{a}_{\bf k}(t) + \tilde{f}_k^*(t) \tilde{a}^\dagger_{-{\bf k}}(t)$, and requires $[\tilde{a}_{\bf k}(t),\tilde{a}^\dagger_{-{\bf k}}(t)] = (2\pi)^d \delta^d({\bf k}-{\bf k}')$ for all $t$, then $(\tilde{f}_k(t),\tilde{f}^*_k(t))$ and $(f_k(t),f^*_k(t))$ are related through the standard Bogoliubov transformation.

As $m^2$ is slowly varying and positive at large negative times, the evolution is adiabatic and we require that the $f_k$ coincide asymptotically with the positive frequency WKB solutions $\tilde{f}_k$ as $t \to -\infty$.  Thus $|0\rangle$ is the ``out'' vacuum in the parlance of \cite{Birrell:1982ix, Kluger1998}, and we assume that our system has been prepared in this vacuum. All expectation values unless otherwise indicated are with respect to this state.

\subsection{The scaling limit}
\label{Sec:QuantumScalingLimit}
In the classical context, we first solved the complete dynamical problem for arbitrary initial condition and protocol choice. We then established that all local physical quantities reduced to universal forms \emph{independent of the details of the protocol away from the critical point and the initial conditions} in the KZ scaling limit. We repeat the same exercise here by using WKB methods to solve Eq.~\eqref{Eq:ModeEquation} for a general TCP/CCP in $m^2(t;\tq)$.

Assume that the system is initially prepared in the vacuum state. The solution to Eq.~\eqref{Eq:ModeEquation} is well-approximated by the positive oscillatory WKB solution $\tilde f_k$ at early times, and by a linear combination of the WKB solutions, $\tilde f_k$ and $\tilde f^*_k$, at late times, as long as the frequency, $1/\Omega_k$, is much larger than the rate at which the frequency changes, $|d\log\Omega_k/dt|$. This yields the condition, $|dm^{-1}/dt| \ll 1$. Within the window $|t| < \tq$, we use the leading order expansion for $m^2(t;\tq)$ in $t/\tq$,
\begin{equation}
\label{Eq:msquareExpansion}
m^2(t;\tq) = m_0^2 \left( -{t \over \tq} \right)^a
\left[ 1 + a_1 {t \over \tq} + \ldots \right] \,,
\end{equation}
to conclude that the WKB solutions are valid as long as $|t| \gg \tk$ or $|\scale{t}| \gg 1$. Note that $\tk$ using Eq.~\eqref{Eq:KZtime} is,
\begin{equation}
\tk = \tq^{a \over 2+a} / m_0^{2 \over 2+a} \,.
\end{equation}
Directly solving the mode equation using the expansion in Eq.~\eqref{Eq:msquareExpansion} yields a solution valid in the region $|t|\ll \tq$ or $|\scale{t}| \ll \tq/\tk$. Thus, the overlap in the ranges of the validity of the direct solution and the WKB one is:
\begin{equation}\label{MatchingRegion}
1 \ll \scale{t} \ll {\tq \over \tk} \,.
\end{equation}

We are now in a position to take the KZ scaling limit whereby $\tq\rightarrow \infty$ and all quantities are measured in units of the KZ length and time. We first notice that the region of overlap in Eq.~\eqref{MatchingRegion} diverges. The mode equation near $t=0$ also simplifies in scaled units. We see from Eq.~\eqref{eq:Wron} that $f_k$ carries dimensions of $\sqrt{t}$, so its scaling form is
\begin{equation}
f_k \sim \sqrt{\tk} \scale{f}_{\scale{k}} \,,
\end{equation}
where $\scale{k} = \lk k$ as usual and $\lk=\tk$. The scaled mode equation in the window $|t|<\tq$ before any limits are taken is
\begin{equation}\label{Eq:SME}
\left[ {d^2 \over d\scale{t}^2} + \scale\Omega_{\scale{k}}^2(\scale{t}) \right] \scale{f}_{\scale{k}} = 0 \,,
\end{equation}
where $\scale\Omega_k^2 = \scale{k}^2 + \scale{m}^2(\scale{t})$ and
\begin{equation}\label{Eq:mhatForm}
\scale{m}^2(\scale{t}) = (-\scale{t})^a \left[ 1 + a_1 {\tk \scale{t} \over \tq} +
  a_2 \left( {\tk \scale{t} \over \tq} \right)^2 + \ldots \right] \,.
\end{equation}
In the KZ scaling limit, $\tk/\tq \rightarrow 0$ and the corrections in the square bracket vanish, resulting in the simpler scaled mode equation,
\begin{equation}\label{Eq:ScaledModeEquation}
\left[ {d^2 \over d\scale{t}^2} + \scale{k}^2 + (-\scale{t})^a \right] \scale{f}_{\scale{k}} = 0 \,.
\end{equation}
The goodness of the approximation is parametrically controlled by the smallness of the parameter $\tk/\tq$. Corrections to the diverging overlap in the ranges of the validity of the solution to the scaled equation above and the WKB one is also controlled by the same small parameter $\tk/\tq$. Thus, the solution to Eq.~\eqref{Eq:ScaledModeEquation} picked out by starting with a positive frequency WKB solution at early times is the same as the one picked out by applying the early-time positive frequency condition directly to the solutions of the limiting equation, Eq.~\eqref{Eq:ScaledModeEquation}.  This is the sense in which the mode functions in the scaling limit are universal.  It is worth noting that these considerations are merely an elaboration of the standard arguments used to justify turning point formulas in standard WKB treatments of the time-independent Schr\"odinger equation in the geometric optics limit.

Finally, let us note that the above considerations will not apply to the marginal ECP. For a given non-universal regularization of the protocol at small time ($t<\tq$), the direct solution to the mode equation in the power-law regime ($t\gg \tq$) has no overlap with the positive frequency WKB solution at early times. As we discuss later, this implies that some information about the short time regularization {\it must} enter the scaling limit.

\subsection{Quasi-particles, heat density and diagonal entropy}
As the Gaussian problem is integrable, quasi-particles are well-defined and infinitely long-lived. The quasi-particle number at momentum $\bf{k}$, $\mathcal{N}_k$, is defined as the expectation value of the occupation of mode $\bf k$ and has the scaling form,
\begin{align}
\mathcal{N}_{\scale{k}} (\scale{t}) &\equiv \langle  \tilde{a}^\dagger_{\bf k}(t) \tilde{a}_{\bf k}(t) \rangle \nonumber \\
& =
 \frac{1}{2\vert \scale{\Omega}_{\scale{k}}\vert} \vert \partial_{\scale{t}} \scale{f}_{\scale{k}}+i\scale{\Omega}_{\scale{k}} \scale{f}_{\scale{k}} \vert^2 \,.
\end{align}
The excess energy density is given by $q(t;\tq) = \int d^d k/(2\pi)^d \Omega_{k} \mathcal{N}_k$ and is easily seen to obey the scaling form conjectured in Eq.~\eqref{Eq:esscalingform}. A definition of the entropy density of the system at each instant of time is through the diagonal entropy density. The diagonal entropy density, $s$, is the entropy density of the diagonal components of the density matrix $\rho(t;\tq)$ in the many-body adiabatic basis. In the Gaussian problem, $s$ is additive in the label $\vct{k}$ and this definition simplifies to,
\begin{align*}
s(t;\tq) =   - \int \frac{d^dk}{(2\pi)^d} \sum_{m} \rho_{m\vct{k},m\vct{k}}\log(\rho_{m\vct{k},m\vct{k}})\,.
\end{align*}
Here $\rho_{m\vct{k},m\vct{k}}$ is the absolute value squared of the overlap between the time-evolved wave-function $|\psi(t)\rangle$, and the $m_{th}$ excited state of the harmonic oscillator labeled by $\vct{k}$. The integral only runs over half the volume in $k$-space as the modes, $\vct{k}$ and $-\vct{k}$, are coupled by the Hamiltonian. $\rho_{m\vct{k},m\vct{k}}$ can be directly computed in the Schr\"odinger picture because the time-evolved wavefunction is known in terms of the mode functions. In the eigenbasis of the operator $\phi_{\vct{k}}$ defined in Eq.~\eqref{Eq:phikdefinition}, the wavefunction is a Gaussian \cite{Dantas:1992aa},
\begin{align}
\langle \phi_k | \psi \rangle \propto  \exp\left[-\int \frac{d^dk}{(2\pi)^d}   \left(\frac{1}{|f_k(t)|^2} - i \frac{\dot f_k(t)}{f_k(t)}\right) \phi_{\vct{k}}\phi_{-\vct{k}}\right] \,\nonumber
\end{align}
up to normalization and time-dependent phases. In the scaling limit, $f_k(t)\sim \sqrt{\tk}  f_{\scale{k}}(\scale{t})$ and $s$ has the scaling form predicted in Eq.~\eqref{Eq:esscalingform}. As all that is at issue is the scaling of the mode functions $f_k(t)$, the scaling forms of correlation functions may also be easily verified.
\begin{align}\label{Eq:GreenFromMode}
G_{\phi\phi}(k,t;\tq) \equiv \vert f_{k}(t)\vert ^2\sim \tk \vert \scale{f}_{\scale{k}}(\scale{t})\vert ^2 \sim \tk \mathcal{G}_{\phi\phi}(\scale{k},\scale{t}) \,.
\end{align}


\subsection{Linear protocol}

\noindent We now turn to a particular protocol, namely the linear quench, where the KZ time is $\tk=(\tq/m_0^2)^{1/3}$.
For this case the mode equation can be solved in closed form to give

\begin{align} \label{Eq:AiryChoice}
\scale{f}_{\scale{k}}({\scale{t}}) =
             \sqrt{\frac{\pi}{2}}\left[ \text{Bi}({\scale{t}}-{\scale{k}}^2) + i\, \text{Ai}({\scale{t}} - {\scale{k}}^2)\right] \,,
\end{align}
where Ai and Bi are Airy functions of the first and second kind. The scaling function of the two-point function from Eq.~\eqref{Eq:GreenFromMode} is:
\begin{align}
\mathcal{G}_{\phi\phi}(\scale{k},\scale{t}) = \frac{\pi}{2} \left( \text{Ai}^2({\scale{t}}-{\scale{k}}^2) + \, \text{Bi}^2({\scale{t}} - {\scale{k}}^2) \right) \,.
\label{eq:G_quantum_TCP}
\end{align}
As a check we may retrieve the equilibrium result when $\hat{t} \rightarrow - \infty$ holding $k\xi$ or $\scale{k}/\sqrt{-\scale{t}}$ fixed.
\begin{align}
\mathcal{G}_{\phi\phi}(\scale{k}, \scale{t}) \sim  \frac{1}{2\scale{\Omega}_{\scale{k}}} \,.
\end{align}
The equilibrium result is not retrieved when $\scale{t}\rightarrow \infty$ because of the pathology of the inverted $\phi^2$ term when $\scale{t}>0$. Instead, $\mathcal{G}_{\phi\phi}$ grows exponentially with $\scale{t}$. From $G_{\phi\phi}$ we can calculate the non-equilibrium correlation length,
\begin{align}
\xi_{\rm ne}\sim \tk  \sqrt{\frac{\partial_{\scale{t}} (\text{Ai}(\scale{t})^2 + \text{Bi}(\scale{t})^2)} {\text{Ai}(\scale{t})^2 + \text{Bi}(\scale{t})^2}}\,.
\label{eq:xi_sqr_quantum_TCP}
\end{align}
whose scaling form $\mathcal{L}_{\rm ne}$ is plotted in Fig.~\ref{fig:xi_sqr_quantum_TCP}(a). At large positive $t$, $\mathcal{L}_{\rm ne}$ grows polynomially instead of exponentially with time, despite the instability of the adiabatic theory with an inverted $\phi^2$ term.
\begin{figure}[!]
  \centering

        {\includegraphics[clip,width=0.95\hsize]{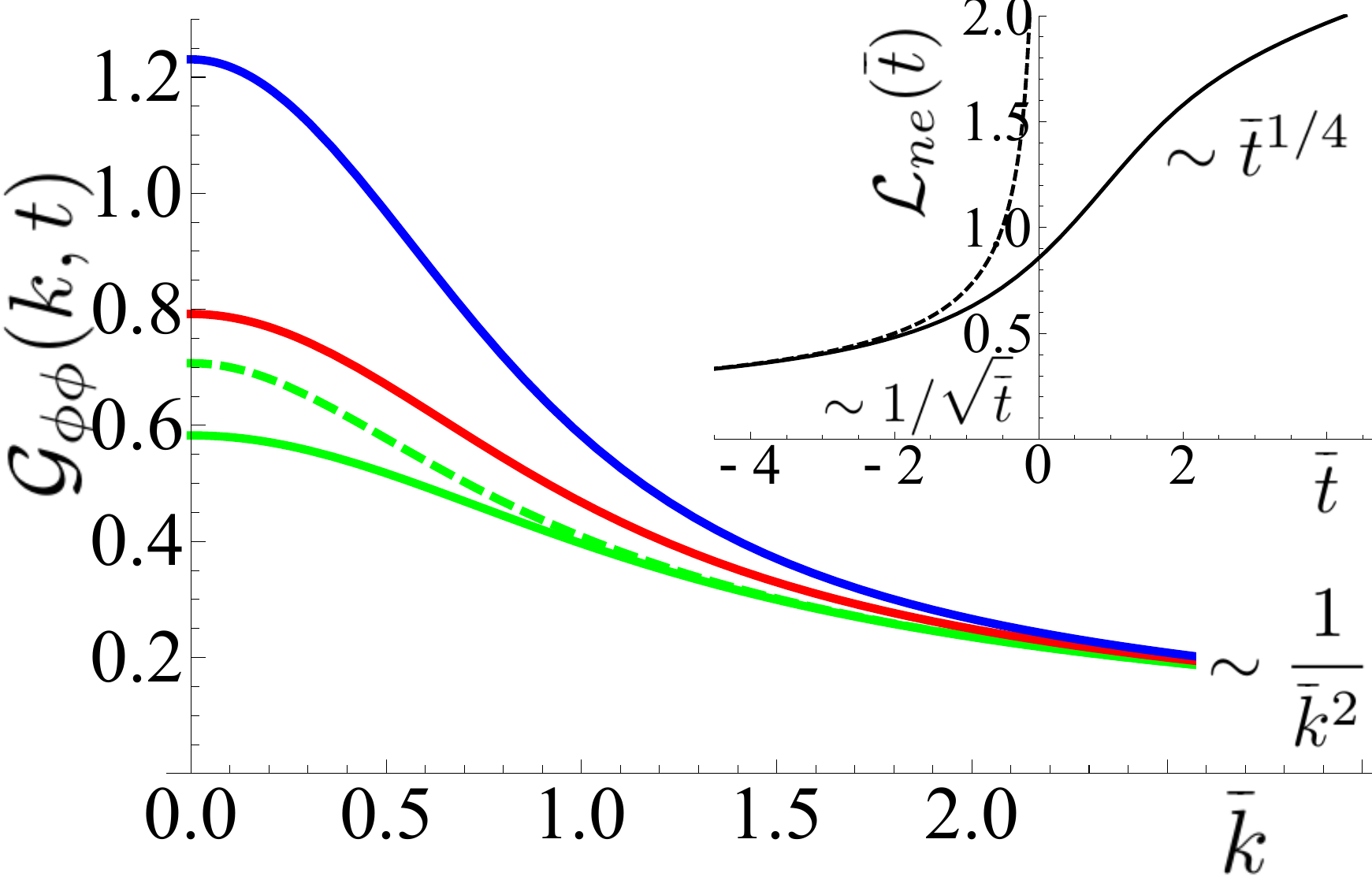}}
  \caption{ $G_{\phi\phi}$ for the linear TCP. The blue, red and green solid lines are respectively at $\scale{t}=0.5,0$ and $-0.5$. The green dashed line is the correlator if the system were in equilibrium at $\scale{t}=-0.5$. Inset: $\mathcal{L}_{\rm ne}$ vs $\scale{t}$ (solid) and $\xi/\lk$ (dashed).}
  \label{fig:xi_sqr_quantum_TCP}
\end{figure}
As the adiabatic problem is pathological for $\scale{t}>0$, we can only sensibly talk of quasiparticle occupations and thermodynamic quantities as
long as $\scale{t} \le 0$. In Fig.~\ref{fig:quasiparticle_number_linear}, we plot $\mathcal{N}_{\scale{k}}({\scale{t}})$ and $\mathcal{S}_{\scale{k}}({\scale{t}})$ for various values of ${\scale{k}}$ (the behavior of the excess energy density can be inferred from the number of quasiparticle excitations).  The contribution to the total quasi-particle density from the high-wavenumber ($\scale{k}\gg\sqrt{-\scale{t}}$) modes for $\scale{t} \ll -1$ is finite only if $d<6$:
\begin{align}
\int_{ \sqrt{-\scale{t}} }^{\infty} \frac{d^d\scale{k}}{(2\pi)^d} \mathcal{N}_{\scale{k}}({\scale{t}}) \sim \frac{1}{(-\scale{t})^{3-d/2}} \,\,\,\,\, \textrm{if }d < 6\nonumber.
\end{align}
The total quasi-particle density, and consequently the energy and entropy density, are ill-defined in the scaling limit within this integrable model when $d\geq 6$. Additionally, observe that $\mathcal{N}_{\scale{0}}$ diverges at
${\scale{t}}=0$ because the gap, $\Omega_0$, between the ground and excited state of the oscillator at $\scale{k}=0$ closes at $\scale{t}=0$.

\begin{figure}[htbp]
  \begin{center}
         {\includegraphics[clip,width=0.5\hsize]{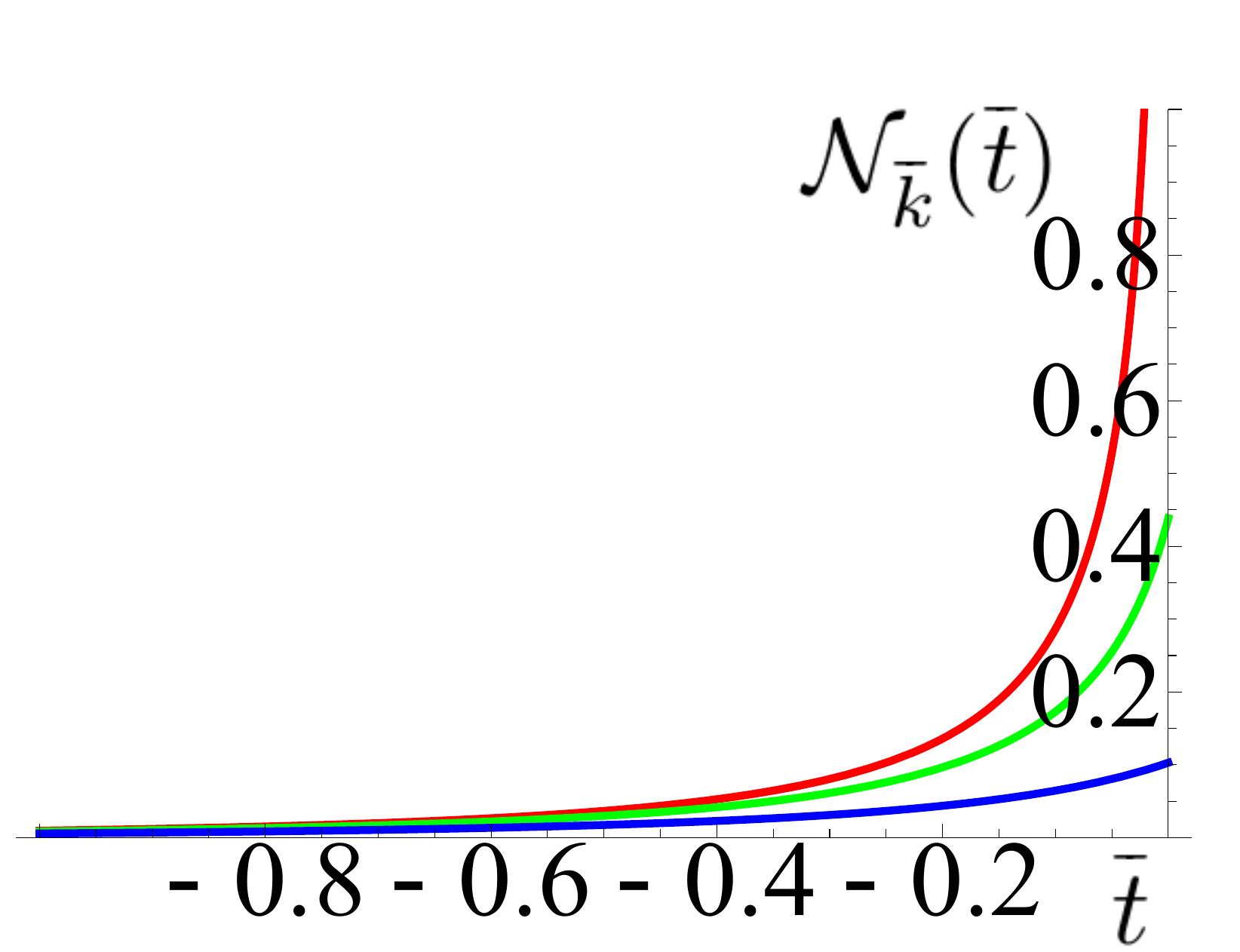}}
            {\includegraphics[clip,width=0.46\hsize]{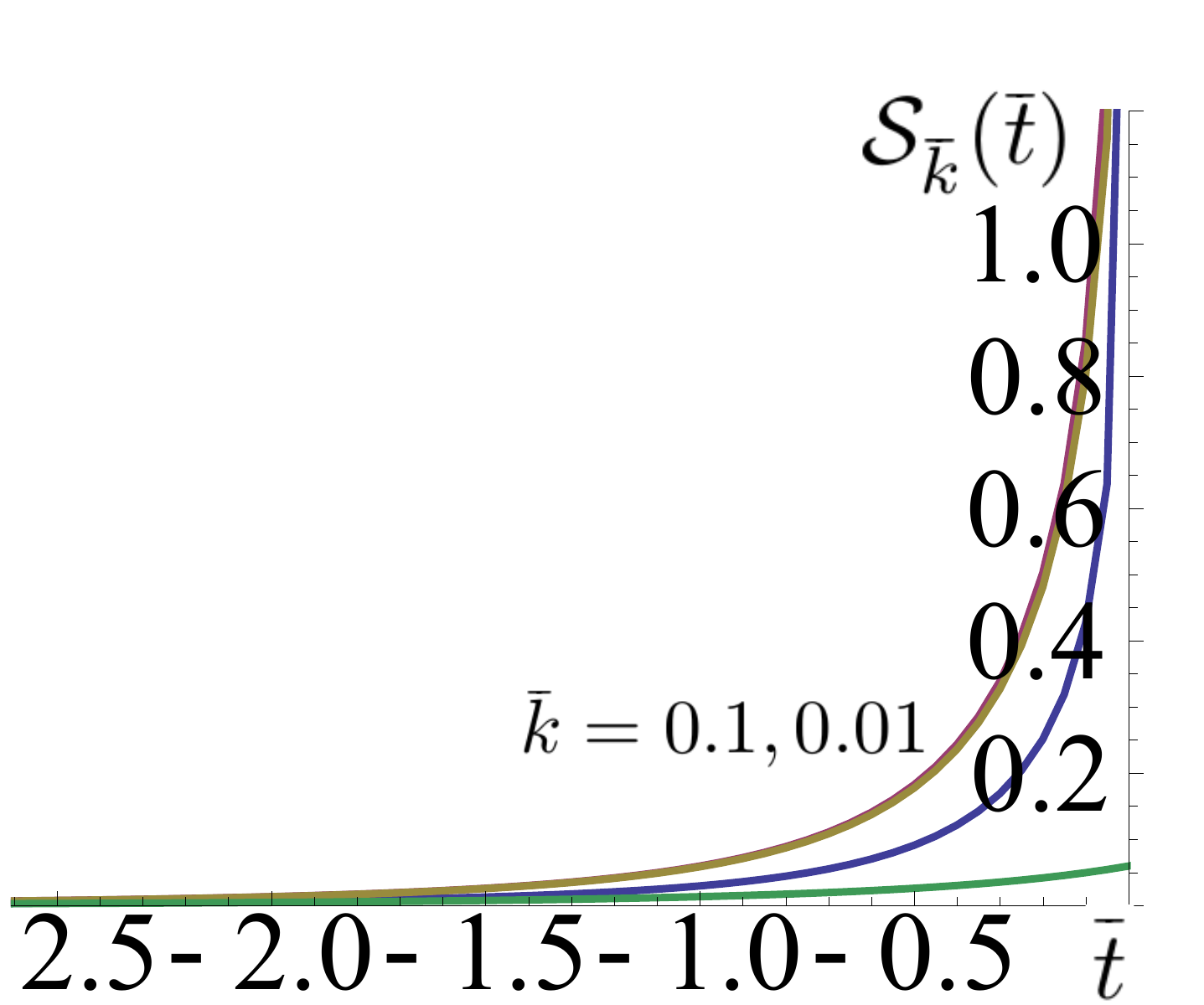}}

  \caption{Left: Quasiparticle number $\mathcal{N}_{\scale{k}}(\scale{t})$ for ${\scale{k}}=0,0.25,0.5$. Right: Diagonal entropy $\mathcal{S}_{\scale{k}}$ for $\scale{k}=0,0.01,0.1,1$ from top to bottom in the linear TCP.}
  \label{fig:quasiparticle_number_linear}
  \end{center}
\end{figure}

\subsection{CCP - Quadratic protocol}
We now turn to a case where the adiabatic problem is well-defined for all $\scale{t}$ -- the quadratic CCP. The KZ time here is $\tk=(\tq/m_0)^{1/2}$. The scaling forms of the mode functions and the equal-time two point function are :
\begin{align}
\scale{f}_{\scale{k}}({\scale{t}}) &=
             2^{-1/4} e^{-\frac{\pi {\scale{k}}^2}{8}} D_{\frac{-1+i{\scale{k}}^2}{2}}(-\sqrt{2} e^{-\frac{i\pi}{4}}{\scale{t}}) \nonumber \\
\mathcal{G}_{\phi\phi}(\scale{k}, \scale{t}) & = \frac{e^{-\frac{\pi \scale{k}^2}{4}}}{\sqrt{2}} \vert D_{\frac{-1+i{\scale{k}}^2}{2}}(-\sqrt{2} e^{-\frac{i\pi}{4}}{\scale{t}}) \vert^2  .
\end{align}
$D_{\nu}(z)$ is the parabolic cylinder function.

As $\scale{t}\rightarrow \pm \infty$ holding $k\xi$ or $\scale{k}/\scale{t}$ fixed, we recover the equilibrium forms,
\begin{align}
\mathcal{G}_{\phi\phi}(\scale{k}, \scale{t}) \sim  \frac{1}{2\scale{\Omega}_{\scale{k}}} \,.
\end{align}
The retrieval of the equilibrium form as $\scale{t}\rightarrow\infty$ is by no means guaranteed. Recent work \cite{Kolodrubetz:2011fj} suggests that this question is intimately tied to the de-phasing of the off-diagonal terms in the time-averaged density matrix in the instantaneous eigenbasis. If the off-diagonal terms do not de-phase, which would be the case for example for the quartic quench, then equilibrium behavior is not expected as $\scale{t}\rightarrow \infty$.

\begin{figure}[htbp]

   {\includegraphics[clip,width=0.5\hsize]{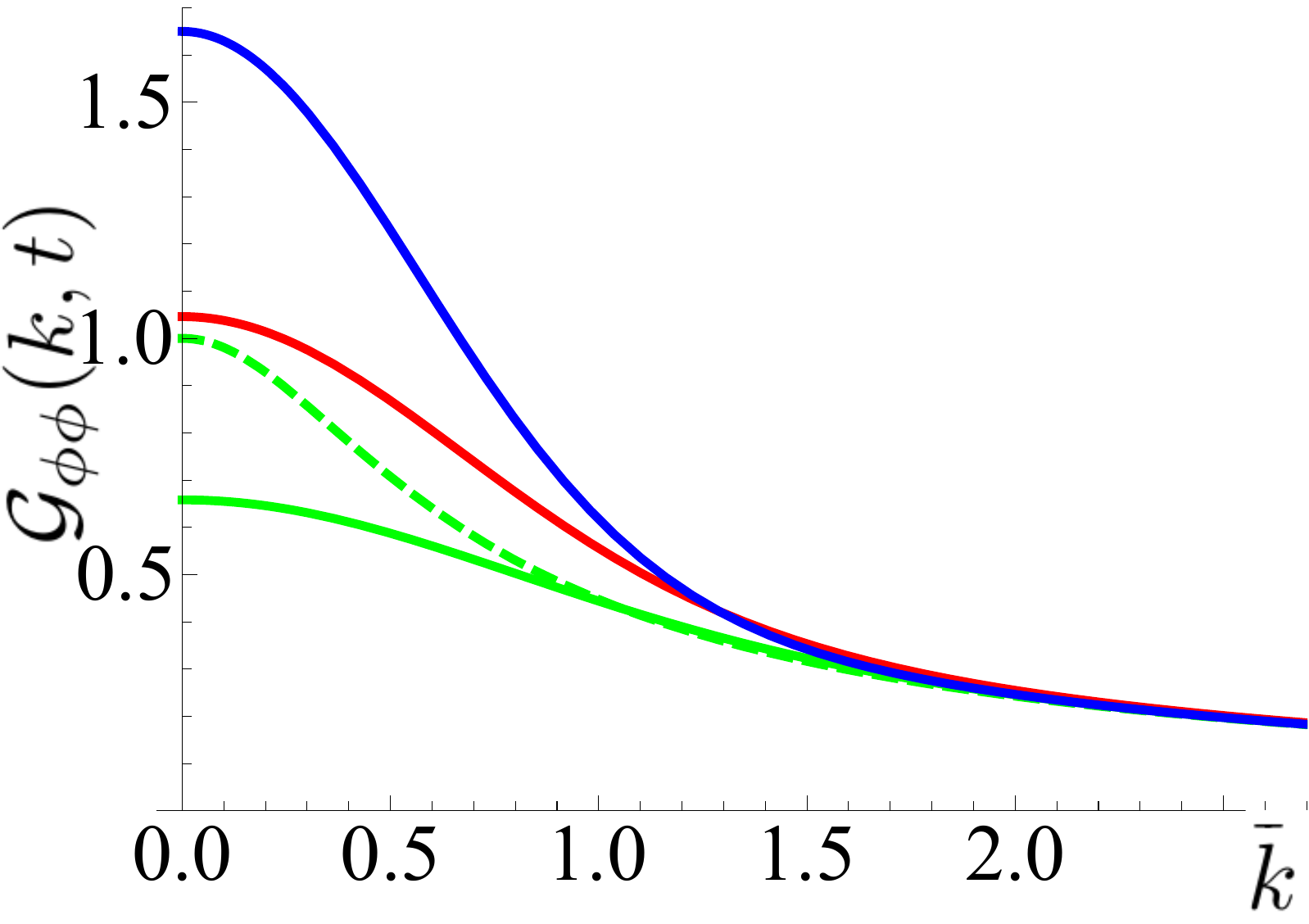}}
   {\includegraphics[clip,width=0.46\hsize]{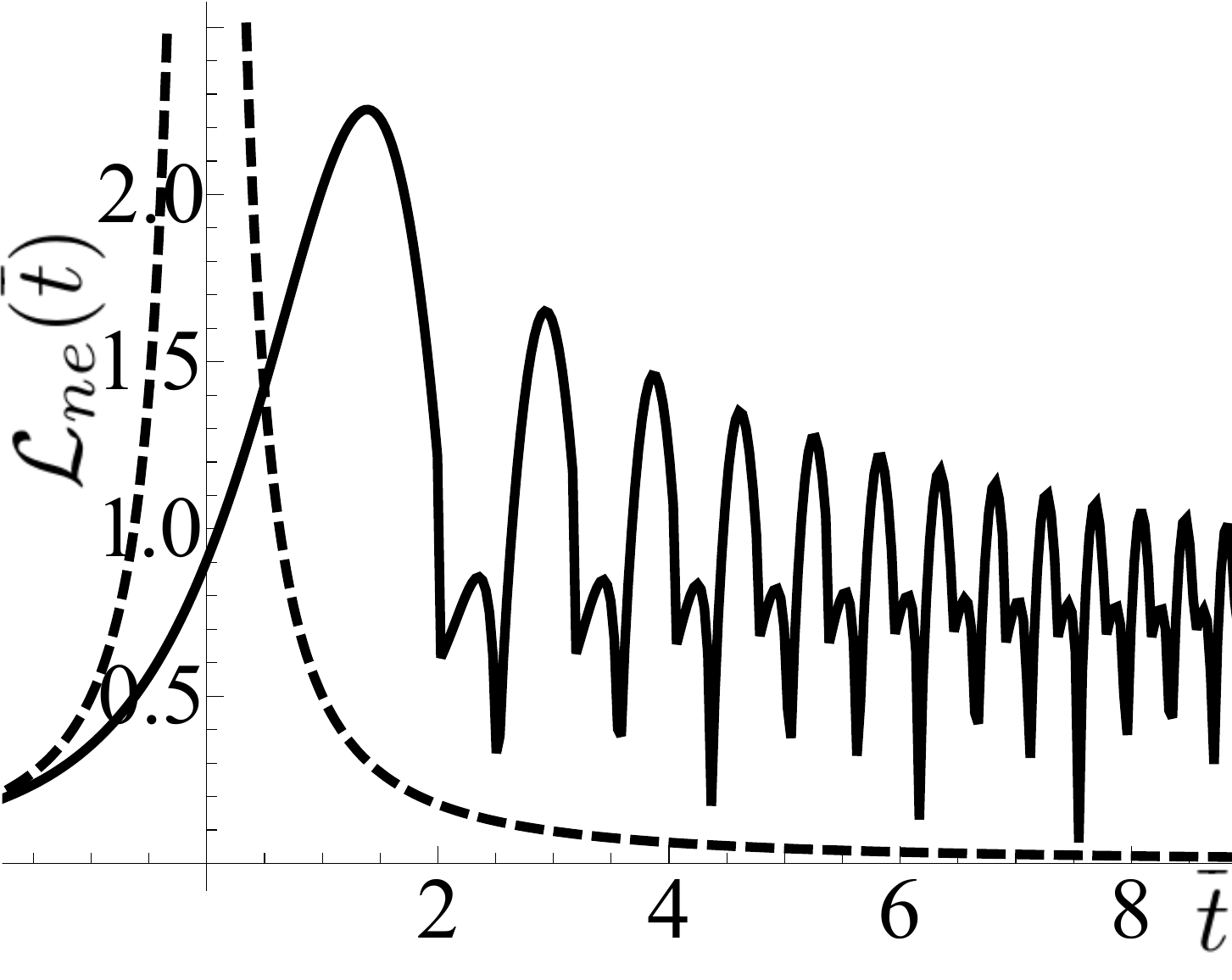}}
   \caption{Left: $G_{\phi\phi}$ for the quadratic CCP. The blue, red and green solid lines are respectively at $\scale{t}=0.5,0$ and $-0.5$. The green dashed line is the correlator if the system were in equilibrium at $\scale{t}=-0.5$. Right: $\mathcal{L}_{\rm ne}$ vs $\scale{t}$ (solid) and $\xi/\lk$ (dashed).}
  \label{Fig:Corr_quadCCP}
\end{figure}

The Gaussian theory imposes further structure on correlators because each momentum mode evolves independently of the others. Consider the time-evolution of the wavefunction of the harmonic oscillator labelled by $\scale{k}$ in the Schr\"odinger picture. Expanding it in the eigenbasis of the Hamiltonian at time $\scale{t}$ and suppressing the label $\scale{k}$,
\begin{equation*}
|\psi\rangle \approx A_0 |0(\scale{t})\rangle + A_1|1(\scale{t})\rangle \ldots\,.
\end{equation*}
The time-evolution is adiabatic when $\scale{t} \gg 1$ and each $|A_i|$ approaches a constant. The relative phase between $A_{i+1}$ and $A_i$ on the other hand grows as :  $\int^{\scale{t}} \scale{\Omega}_{\scale{k}}(\scale{t'}) dt' \sim \scale{t}^2$. This is the origin of the oscillations of period $1/\scale{t}$ at large positive $\scale{t}$ at each $\scale{k}$, seen for example in $\mathcal{G}_{\phi\phi}(\scale{k},\scale{t})$ in Fig.~\ref{Fig:Corr_quadCCP}(a).

The simple definition of $\xi_{\rm ne}$ in Sec.~\ref{Sec:CorrlengthDef} has to be modified as $\mathcal{G}_{\phi\phi}(\scale{k},\scale{t})$ has multiple poles in the complex $\scale{k}$-plane for late times. The pole with the smallest imaginary part determines the decay constant ($1/\xi_{\rm ne}$) over the longest length scales in the two-point function in real-space. The non-equilibrium correlation length is plotted in Fig.~\ref{Fig:Corr_quadCCP}(b). As $\scale{t}\rightarrow\pm \infty$, the envelope of $\xi_{\rm ne}$ behaves as,
\begin{equation*}
\xi_{\rm ne}(\scale{t}) \sim \frac{1}{|\scale{t}|}.
\end{equation*}
This confirms the prediction in Eq.~\eqref{Eq:xineasymptotes} for CCPs with no coarsening physics. For the reason discussed above, it also exhibits characteristic oscillations of period $1/\scale{t}$ at long times.

\begin{figure}[htbp]
\begin{center}
\includegraphics[width=4cm]{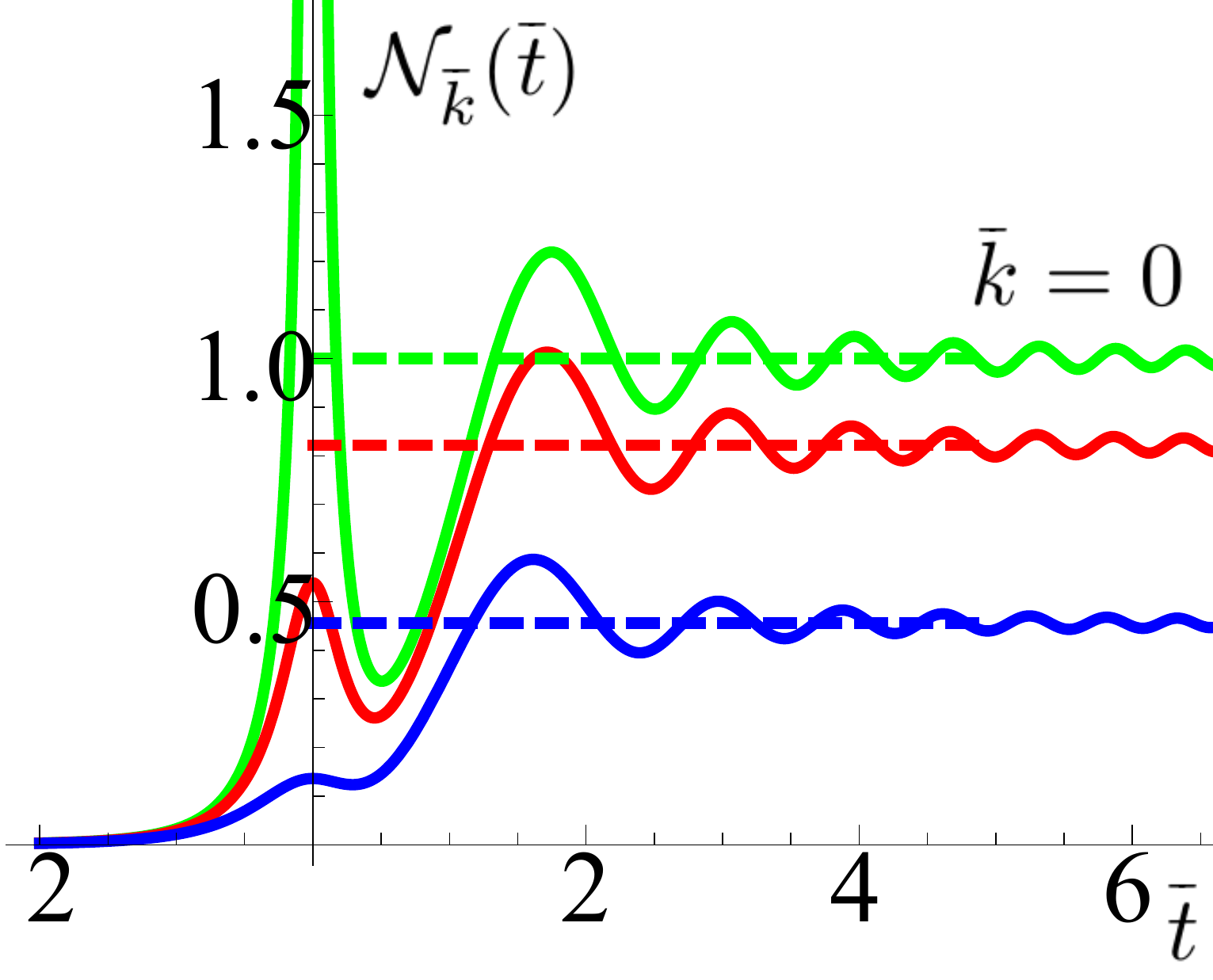}
\includegraphics[width=4cm]{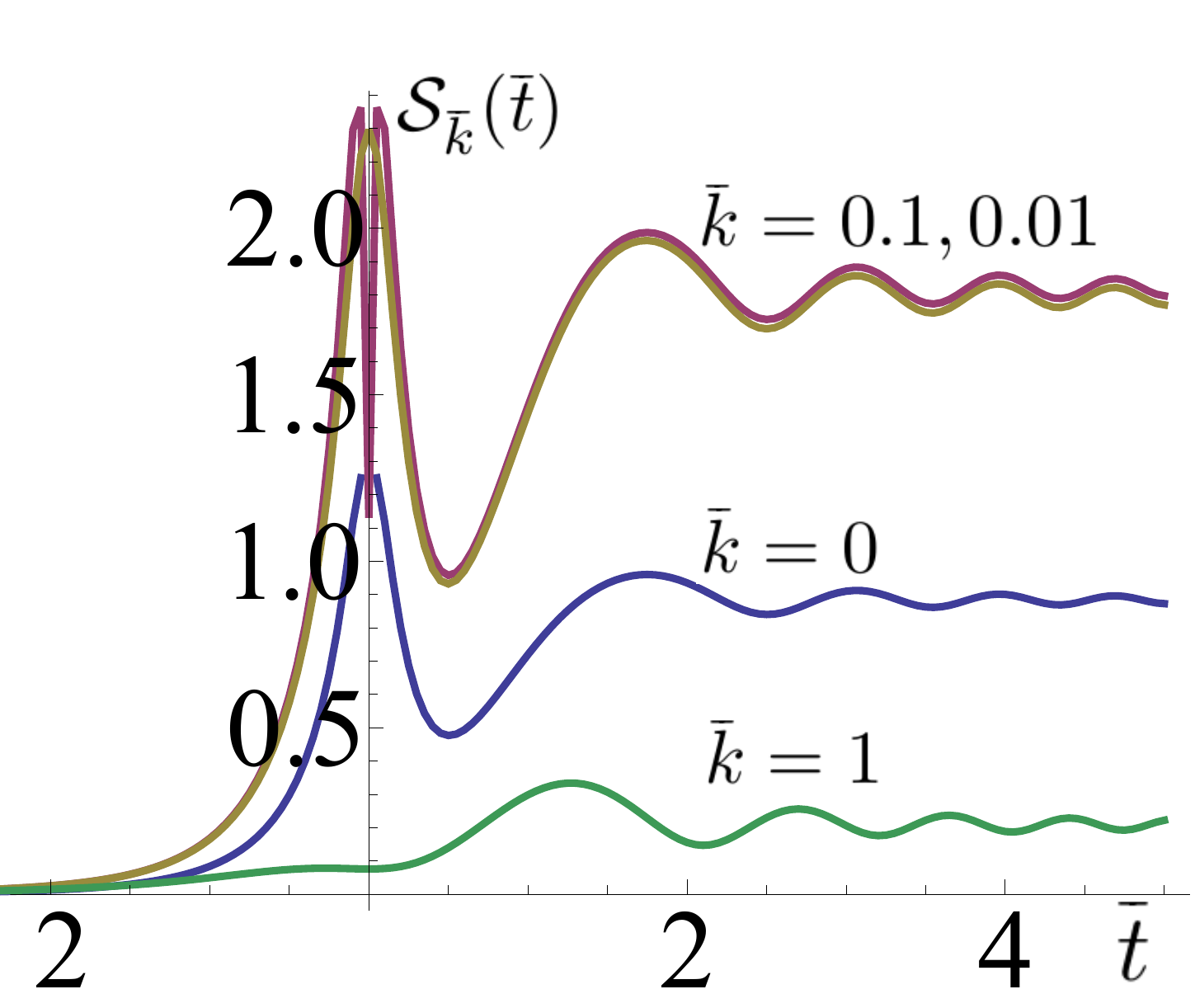}
  \caption{Left: Quasiparticle number $\mathcal{N}_{\scale{k}}(\scale{t})$ for ${\scale{k}}=0,0.25,0.5$. Right: Diagonal entropy $\mathcal{S}_{\scale{k}}$ for $\scale{k}=0,0.01,0.1,1$ from top to bottom in the quadratic CCP.}
  \label{Fig:thermodynamic_quadCCP}
\end{center}
\end{figure}

Finally, we plot $\mathcal{N}_{\scale{k}}(\scale{t})$ and $\mathcal{S}_{\scale{k}}(\scale{t})$ for various values of $\scale{k}$ in Fig.~\ref{Fig:thermodynamic_quadCCP}. In each oscillator, the quasiparticles and entropy is essentially produced in the time-interval $|\scale{t}|\lesssim1$. At late times, both quantities settle to a constant dependent on $\scale{k}$. For example,
\begin{equation*}
\mathcal{N}_{\scale{k}}(\scale{t}) \rightarrow e^{-\pi \scale{k}^2} \textrm{ as } \scale{t}\rightarrow \infty.
\end{equation*}
As was the case for the linear quench, the total quasiparticle and entropy density is finite only if $d<6$.

\subsection{The Marginal ECP}
\label{Sec:GaussianECP}
The relativistic Gaussian theory provides an ideal setting to study the new non-equilibrium states that arise in the marginal ECP.  The protocol is defined by Eq.~\eqref{Eq:mecp} -- $m(t)$ smoothly transitions from a constant to $\theta/t$ over the time-scale $\tq$ and $\dot{\xi_t}$, the parameter that controls adiabaticity, equals $1/\theta$ for $t\gg \tq$. Our naive expectation is that the mode function is of a universal scaling form, $f_k(t)=\sqrt{t}\scale{f}(kt)$, in the scaling limit $t \rightarrow \infty, k \rightarrow 0$ with $kt$ held fixed.

This expectation is violated on two fronts even in the Gaussian problem. The first is that the marginal ECP is by construction unable to entire ``forget'' its early time regularization.  More precisely, observe that our choice of protocols
implies that  $|dm^{-1}/dt| \ll 1$ when $t \ll \tq$ while $|dm^{-1}/dt| = 1/\theta$ when
$t \gg \tq$.  The latter result shows that for small enough $k$, it is not possible to reach the power law regime while remaining adiabatic, in contrast to our discussion of the TCP and CCP cases; thus we should expect that {\it some} non-universal information must make its way into the putative scaling regime.  The second and more striking violation is that of scaling. When $\theta >1/2$ and $|dm^{-1}/dt|$ is closer to obeying the condition of adiabaticity, scaling is violated ``mildly'' and all physical quantities are periodic functions of $\log(t)$. When $\theta =1/2$, scaling is violated logarithmically. The most dramatic violation of scaling is when $0<\theta<1/2$. Here the adiabaticity condition is strongly broken and the scalar field acquires an anomalous dimension.

Let us now describe how these features emerge in the long time, small momentum form of
the mode functions. The mode equation for $t\gg\tq$:
\begin{align}
\left( \frac{d^2}{dt^2} + k^2+ \frac{\theta^2}{t^2} \right)f_{k}(t) = 0.
\label{eq:ECP_allmode_eq}
\end{align}
Define $\lambda \equiv \left|\sqrt{\frac{1}{4} - \theta^2}\right|$ and
\begin{align}
h(\lambda) = \begin{cases}
i \lambda , & \mbox{if } \ \theta > 1/2 \\
0, &  \mbox{if } \ \theta = 1/2\\
\lambda, & \mbox{if } \ \theta < 1/2.
\end{cases}
\end{align}
Setting $k=0$, Eq.~\eqref{eq:ECP_allmode_eq} is solved as:
\begin{align}
f_0(t) =
\begin{cases}
\sqrt{t} \left(\frac{2^{-\lambda} u}{\Gamma(1+\lambda)} \,\left(\frac{t}{\tq}\right)^{h(\lambda)} + \frac{-2^\lambda \Gamma(\lambda)v}{\pi} \, \left(\frac{t}{\tq}\right)^{- h(\lambda)}\right) , \\
\hspace{145pt}\textrm{if } \theta\ne 1/2 \nonumber \\
\sqrt{t}\,( \,u + v \log{\left(\frac{t}{\tq}\right)} \,),  \,\, \mbox{if } \theta = 1/2 \ .
\end{cases}
\end{align}
The units of $t$ are chosen such that $f_0(\tq) \sim \sqrt{\tq}$. The particular $\lambda$-dependent pre-factors are chosen to simplify the solution at all $k$. $v$ and $u$ are non-universal constants in the equation above. They depend on the detailed pre-asymptotic form of the protocol and are fixed by the solution to the
differential equation with the full time dependance of the mass.

Let us now turn to $k \ne 0$. For times greater than $\tq$, the mode equation is solved by the linear combination,
\begin{align}
\label{Eq:ECPExact}
f_k(t) = \sqrt{t} \left[ c_J (k)\, J_{h(\lambda)}(kt) +c_Y(k)\, Y_{h( \lambda)}(kt)
  \right] \,.
\end{align}

$J_\nu(x)$ and $Y_\nu(x)$ are the Bessel functions of order $\nu$ of the first and the second kind. The functions $c_J(k)$ and $c_Y(k)$ are presently undetermined except for the requirement of
smoothness in $k$. However, if we examine the region $t \gg \tq $ at small $kt$, we can
obtain their leading order behavior by requiring that they match smoothly to the $k=0$ forms
presented earlier. This localizes the non-universality to the constants $u,v$ discussed already. For $\theta < 1/2$ and $k$ smaller than some non-universal $k_0$, this leads to the leading behavior:
\begin{align}
c_Y(k) &=  v(k \tq)^\lambda \nonumber \\
c_J(k) &= u(k \tq)^{-\lambda} - v\frac{ (k \tq)^\lambda}{\pi} \cot \pi\lambda \,.
\end{align}
For $\theta > 1/2$, we simply replace $\lambda$ by $i \lambda$. For $\theta = 1/2$, we
find
\begin{align}
\label{Eq:Lambda0Coeff}
c_Y(k) &= {\pi \over 2} v \nonumber \\
c_J(k) &= u   - v \log{k \tq e^\gamma \over 2} \ .
\end{align}
$\gamma$ is the Euler-Mascheroni constant. We emphasize that the mode-function and consequently all physical quantities are well-approximated by Eq.~\eqref{Eq:ECPExact} with the forms of $c_J$ and $c_Y$ above for $t\gg \tq$ and $k\ll k_0$.
\begin{figure}[!]
  \centering
   {\includegraphics[clip,width=0.95\hsize]{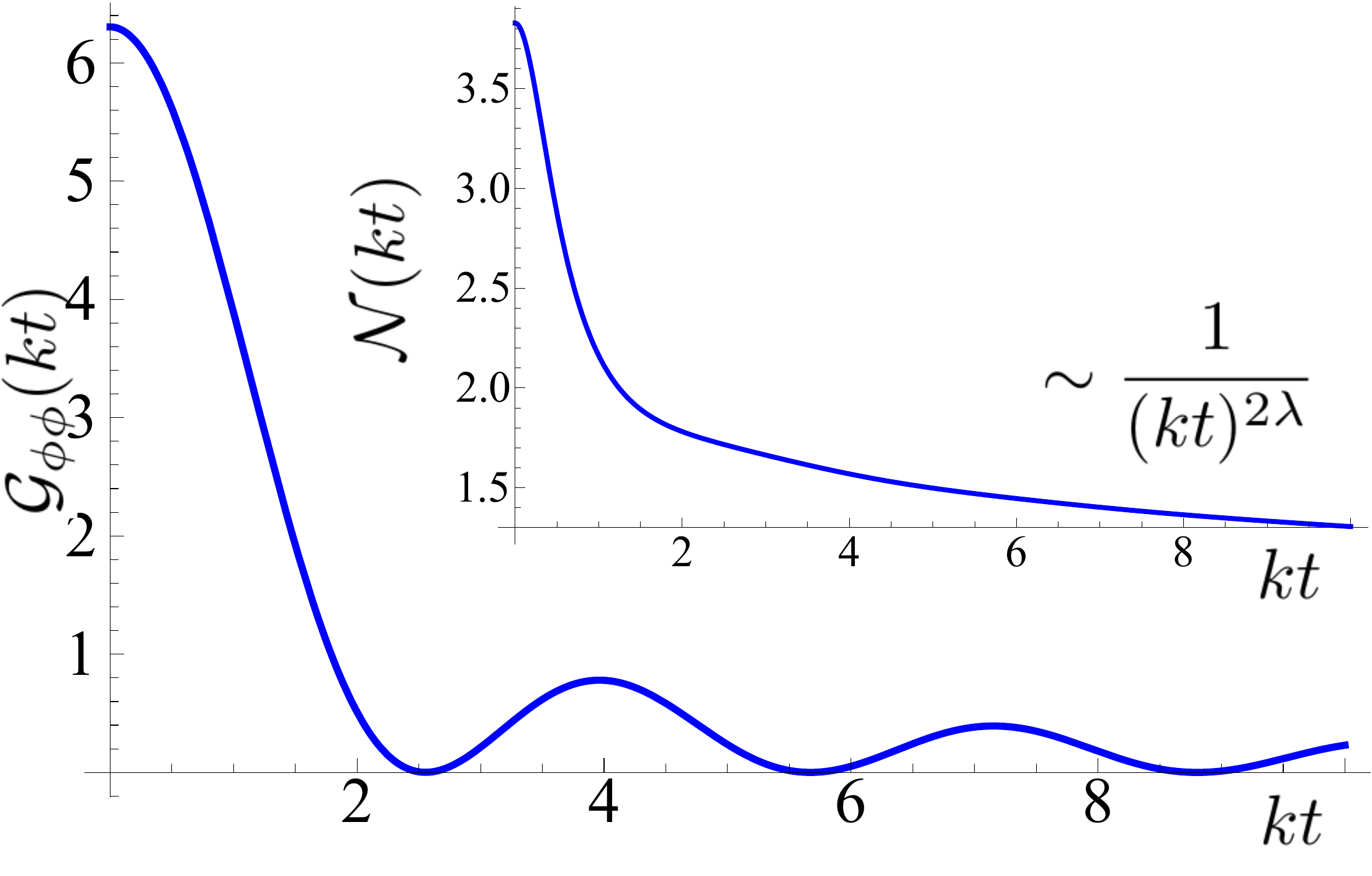}}
 \caption{$\mathcal{G}_{\phi\phi}(kt)$ and $\mathcal{N}(kt)$ vs $kt$ when $\lambda=0.1$ and $u/\tq^\lambda=-2.5i$.  }
  \label{Fig:ecp_lambdareal}
\end{figure}


We now turn to the physical implications of these solutions for an arbitrary choice of $(u,v)$ that satisfy the constraint imposed by the Wronskian condition Eq.~\eqref{eq:Wron}.  First consider the case $0 < \theta < 1/2$, which is strongly non-adiabatic in the sense that $|dm^{-1}/dt| = 1/\theta$ is greater than $2$ and may be large. The mode function in Eq.~\eqref{Eq:ECPExact} can be re-written in the form
\begin{align}
\label{Eq:ModeFunctionLambdaReal}
f_k(t) = t^{1/2} \left(\frac{t}{\tq}\right)^\lambda \left[ g_{+}(kt) + \left(\frac{t}{\tq}\right)^{-2\lambda} g_{-}(kt) \right] \,.
\end{align}
$g_+$ and $g_-$ are linear combinations of the two Bessel functions and involve the constants $u,v$. Observe that in the limit of large $t$ ($t\gg \tq$) with $kt$ fixed, the ratio of the second to the first term in the expression above decreases as $(t/\tq)^{-2\lambda}$ \footnote{The reader may worry about dropping the second term at the zeros of $g_+(kt)$. Fortunately, the correction to $f_k(t)$ due to $g_{-}(kt)$ at these points decreases as $t\rightarrow\infty$.}. Thus, in the scaling limit, $f_k(t)$ simplifies to
\begin{align}
f_k(t) \sim u' t^{1/ 2+\lambda}  \scale{f}(kt) \sim u' t^{1/ 2+\lambda}  \frac{J_{\lambda}(kt)}{(kt)^\lambda}.
\end{align}
$u'=u/\tq^\lambda$ above. The scaling forms predicted in Sec.~\ref{Sec:MarginalECPScalingTheory} \emph{do not hold}. Instead, the form above is the one expected when the field $\phi$ has an anomalous dimension $\lambda$. The scaling form of $G_{\phi\phi}$ with the modified dimension of $\phi$ is
\begin{align}
\label{Eq:GLambdaReal}
G_{\phi\phi}(k,t) \sim t^{1+2\lambda} \mathcal{G}_{\phi\phi}(kt) \,.
\end{align}
In real-space, this implies that the equal-time two-point correlator decays as $1/x^{d-1-2\lambda}$ at fixed $x/t$. Analogously, the quasi-particle number $N_k(t)$ has the scaling form $t^{2\lambda} \mathcal{N}(kt)$. The scaling functions are:
\begin{align*}
\mathcal{G}_{\phi\phi}(kt) &= | u' |^2\scale{f}^2 \\
\mathcal{N}(kt) &= |u'|^2 \frac{|(kt) \partial_{kt} \scale{f} + i \sqrt{(kt)^2 + (\theta)^2} \scale{f}|^2}{2\sqrt{(kt)^2 + (\theta)^2}}.
\end{align*}
They are shown in Fig.~\ref{Fig:ecp_lambdareal}. Three comments are in order. First, in the marginal ECP, $\xi\sim t$ and the time-scale for a change in $m$ is the same as $\xi=\xi_t$. We therefore expect that the non-equilibrium correlation length $\xi_{\rm ne}$ also grow linearly in time. This is indeed the case; the oscillations in $\mathcal{G}_{\phi\phi}(kt)\sim t$ in Fig.~\ref{Fig:ecp_lambdareal} are of order one period and indicate a peak at $r\sim t$ in the real-space correlator $\mathcal{G}_{\phi\phi}(r/t)$. Second, the excess energy density above the instantaneous vacuum \emph{decreases} as $1/t^{1-2\lambda}$. The marginal ECP thus leads the system to the critical point through a family of new, non-equilibrium states. Finally, all scaling functions are known up to a multiplicative constant ($u'$) in the scaling limit.

\begin{figure}[htbp]
\begin{center}
 \includegraphics[clip,width=0.95\hsize]{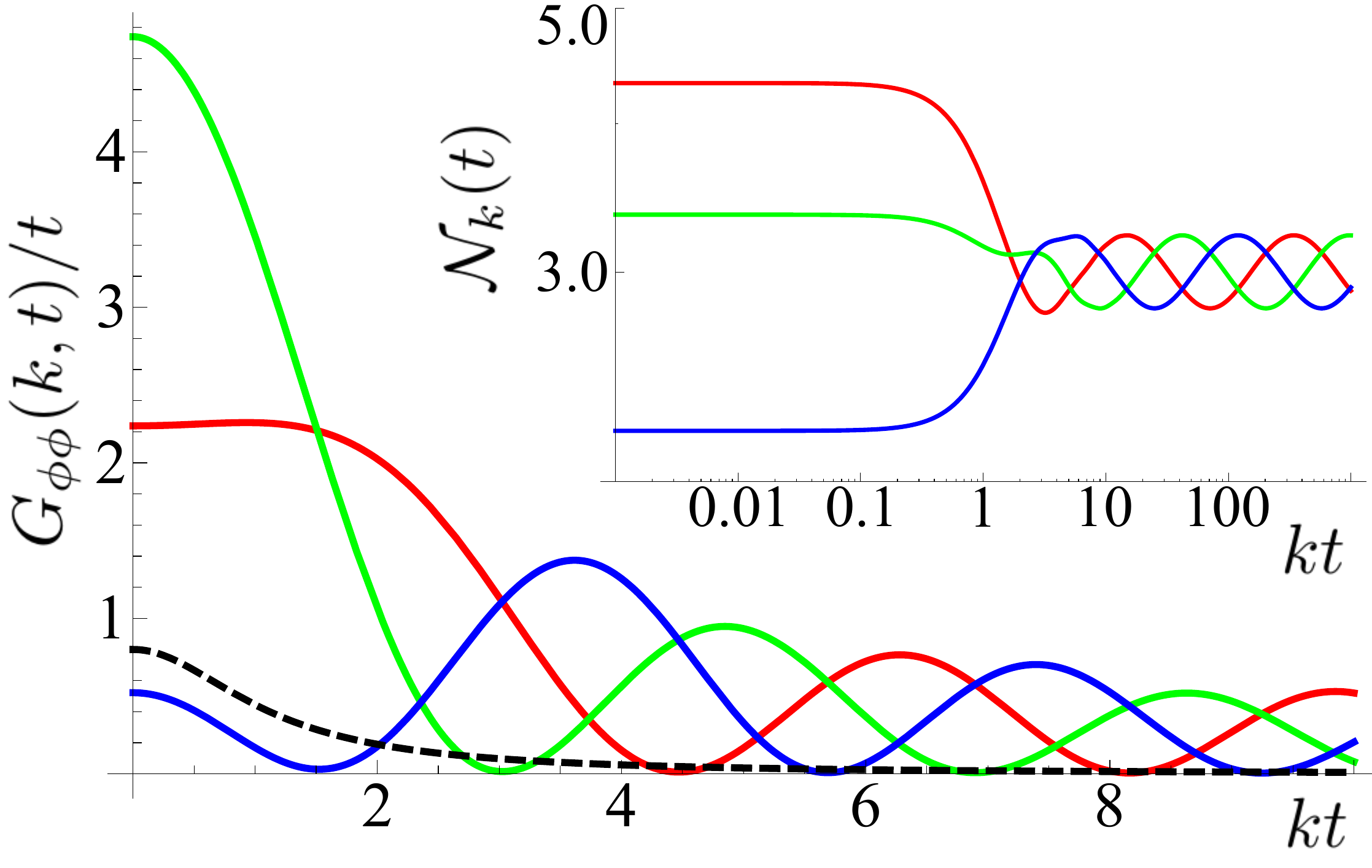}
\label{Fig:ecplambdaimag}
\caption{$G_{\phi\phi}(k,t)/t$ and $\mathcal{N}_k(t)$ vs $kt$ when $\lambda=i$ plotted at three equally spaced time on the $\log(t)$ scale : $t = 10$ (red), $10 \,e^{\pi |\lambda|/3}$ (green) and $ 10\, e^{2\pi|\lambda|/3}$ (blue). The dashed line is the adiabatic response. $(u,v,\tq)$ are chosen to be $( \sqrt{3/2},1,1)$. }
\end{center}
\end{figure}
Let us consider the case $\theta>1/2$, which is weakly non-adiabatic in the sense that $|dm^{-1}/dt| = 1/\theta$ is less than $2$ and may be small.  A re-writing of the mode function informs us that the scaling is violated only by phases:
\begin{align}
f_k(t) =\sqrt{t} \left[ \left(\frac{t}{\tq}\right)^{i|\lambda|} g_{+}(kt) + \left(\frac{t}{\tq}\right)^{-i|\lambda|} g_{-}(kt) \right] \,.
\end{align}
These phases affect other physical quantities in the scaling limit, but in relatively mild ways as compared to the previously seen factors of $t^\lambda$.  For instance, at late times holding $kt$ fixed,
\begin{align}\label{eq:Gperiodic}
G_{\phi\phi}(k,t) \sim t \, \mathcal{G}_{\phi\phi}(kt, e^{i|\lambda| \log (t/\tq)}) \,,
\end{align}
where we have expressed $t^{i|\lambda|} = e^{i|\lambda| \log t}$ in order to emphasize that this quantity is periodic in $\log t$.  The scaling function ${\cal N}_k$ is also periodic in $\log t$. The two ``almost-scaling'' functions ${\cal G}_{\phi\phi}$ and ${\cal N}_k(t)$ are plotted in Fig.~\ref{Fig:ecplambdaimag} for three equally spaced values of $\log t$ within a decade. The late time behavior of $\mathcal{G}_{\phi\phi}$ in the marginal ECP differs markedly from the adiabatic response, show as a dashed line in the plot.  The period oscillations in the two-point correlator imply that $\xi_{\rm ne}\sim t$.  The smaller excess energy density proportional to $1/t$ is an indicator that the system is closer to being in equilibrium than for the strongly non-adiabatic case $0 < \theta < 1/2$.  In a sense, the factors of $t^{i|\lambda|} = e^{i|\lambda| \log t}$ can be regarded as introducing only ``logarithmic'' modifications of scaling.

The case $\theta=1/2$ displays the properties of both the cases discussed above. The relations in Eq.~\eqref{Eq:Lambda0Coeff} imply the form
\begin{align*}
f_k(t) = \sqrt{t} \left[ g_{+}(kt) + \log(t/\tq) g_{-}(kt) \right] \,.
\end{align*}
At late times with $kt$ fixed, the second term dominates the first and the mode function has the scaling form
\begin{align*}
f_k(t) &\sim u\,\sqrt{t} \log{t} \,J_0(kt)
\end{align*}
All scaling forms are thus modified by pre-factors of powers of $\log t$. The logarithmic violation of scaling is similar to the weakly non-adiabatic case $\theta > 1/2$, while the modification of the dimension of $\phi$ is similar to the strongly non-adiabatic case. The quasi-particle number here diverges logarithmically in time.

The excess energy density injected into the system decreases as $t$ increases in the strictly Gaussian theory. The marginal ECPs thus define a family of non-equilibrium states that lead to the critical point that are universal when the Gaussian fixed point is stable. The vanishing excess energy density further suggests that the non-equilibrium states generated by the marginal ECP survive even when the fixed point is not Gaussian.

Finally, let us comment on the ECP in the transverse field Ising model in (1+1) dimensions which is famously
a model of free fermions. Here $\nu=1$ and $z=1$ so the relevant power is now $a=1$. The explicit solution of
the fermionic mode equations exhibits the analogs of our weakly non-adiabatic regime
with corrections to scaling that are periodic in $\log t$ \cite{Chandran:2012aa}. Connecting this behavior to our Gaussian
results discussed above by continuation in the number of dimensions is an interesting challenge for future
work.
\subsubsection{Analogies to $dS_{d+1}/CFT_d$}
We showed above that in the long time limit of the solution to the mode-equation in the strongly non-adiabatic case Eq.~\eqref{Eq:ModeFunctionLambdaReal}, the scalar field acquires an anomalous dimension $\lambda$. If we instead took the opposite limit of small time ($t\sim \tq$), the second term is more important that the first and the anomalous dimension of the scalar field is $-\lambda$. Thus, the effective scaling dimension of $\phi$ in dimension $d$ is
 \begin{align}
  \tilde\Delta_\pm = {d-1 \over 2} \pm \lambda \,. \label{eq:tildeDims}
 \end{align}
 at short and long times respectively. Readers familiar with the $(A)dS_{d+1}/(C)FT_d$ literature, in particular \cite{Strominger:2001pn,Witten:2001ua}, will note a similarity between the result Eq.~\eqref{Eq:GLambdaReal} and the $d$-dimensional field theory Green's function obtained in the presence of a double-trace deformation. We can make the analogy closer by noting that the modified scalar field
 \begin{align*}
  \varphi \equiv t^{{d-1} \over 2} \phi
 \end{align*}
obeys an equation of motion which follows from the Lagrangian
 \begin{align}\label{eq:LdS}
  {\cal L} = {1 \over 2} \sqrt{-\det g_{\alpha\beta}} \left( 
    -g^{\mu\nu} \partial_\mu \varphi \partial_\nu \varphi - M^2 \varphi^2 \right) \,,
 \end{align}
where $g_{\mu\nu}$ is the metric of de Sitter space, $dS_{d+1}$:
 \begin{align}
  ds^2 = {L^2 \over t^2} (-dt^2 + d\vec{x}^2) \,.  \label{eq:dSL}
 \end{align}
The length scale $L$ is arbitrary.  The de Sitter mass must be given by
 \begin{equation}\label{eq:dSmass}
  M^2 L^2 = \theta^2  + {d^2-1 \over 4}
 \end{equation}
in order for the equation of motion from Eq.~\eqref{eq:dSL} to agree with Eq.~\eqref{eq:ECP_allmode_eq}. The dimensions Eq.~\eqref{eq:tildeDims} are closely related to the usual ones in dS/CFT:
 \begin{equation}
  \Delta_\pm = {d \over 2} \pm \sqrt{{d^2 \over 4} - M^2 L^2}
    = \tilde\Delta_\pm + {1 \over 2} \,.
 \end{equation}
The difference arises due to an important distinction between the late-time Green's function Eq.~\eqref{Eq:GLambdaReal} and the $d$-dimensional field theory two-point function.  Examples of the latter could be computed in $dS_{d+1}$ in terms of early time (that is, small $t$) properties of mode functions associated to the ``out'' vacuum: that is, mode functions which are purely positive frequency at late times.  Our computation is the reverse of this, in the sense that we investigate late-time properties of mode functions associated with an ``in'' vacuum.  Late time corresponds to the deep interior of $dS_{d+1}$ (more precisely, it is a corner of the global covering space far from the boundary at $t=0$).  Another difference (of lesser consequence) is that in the normal parlance of dS/CFT \cite{Strominger:2001pn}, time flow would be reversed, so that what we call $t=0$ is the far future while $t \to \infty$ is the far past.

In the window $kt \ll 1$, $f_k(t)$ is $k$-independent and equal to $f_0(t)$:
 \begin{equation}
  t^{{d-1} \over 2} f_k(t) \approx a_\varphi t^{\Delta_-} + b_\varphi t^{\Delta_+} \,,
 \end{equation}
where $b_\varphi , a_\varphi$ are known in terms of $u,v,\tq$, and in particular have a definite ratio. This setup now bears a strong resemblance to double-trace operator deformations in AdS/CFT \cite{Witten:2001ua}.  In AdS/CFT, the relativistic conformal symmetry of the boundary theory is broken by the multi-trace deformation.  The presence of different powers of $x$ in the real-space Green's function dependent on the energy scale $1/\tq$ signals a similar breakdown of scaling.  The same can be said of the weakly non-adiabatic case, but comparisons with a boundary field theory are harder in this case because the field theory would have to be non-unitary, similar to violations of the Breitenlohner-Freedman bound \cite{Breitenlohner:1982jf} in AdS/CFT.  However, the $dS_{d+1}$ formulation (\ref{eq:LdS}) does offer some further intuition regarding the weakly non-adiabatic case: Eq.~\eqref{eq:dSmass} shows that large $\theta$ implies large $M^2 L^2$.  This in turn implies that the Compton wavelength of the massive scalar is much smaller than the Hubble scale of $dS_{d+1}$, which is precisely the condition one needs in order to justify a geometric optics approximation.

\section{Closing Remarks}
Our primary purpose in this paper has been to systematize the universal content of the KZ
problem and to emphasize that all physical quantities give rise to universal scaling
functions that span the entire crossover from equilibrium at early times to the late time state.
We have presented model computations that bear out this logic.
Experiments directed towards observing this broader scaling picture would be highly
desirable. We note that the scaling ideas presented here do not rely on the existence of a local order parameter and generalize straightforwardly to the Rajanti-Hindmarsh mechanism \cite{Hindmarsh:2000uq} as we will discuss elsewhere. 

An obvious challenge is to
extend such computations to more physically realizable problems where the field theories
are not as simple. We will present some results on non-trivial, but not physically realizable
field theories, via the AdS/CFT correspondence elsewhere. Another obvious challenge is to
formulate a renormalization group procedure that makes the universality manifest---beyond
the case of stochastic classical models with the associated functional integral formalism
discussed here.

A byproduct of our analysis has been the identification of an especially interesting ECP
which is able to produce anomalous dimensions already at the Gaussian level, through a mechanism similar to the way anomalous dimensions emerge in (A)dS/CFT.  A deeper
understanding of this phenomenon and its examination in interacting contexts is also a fit
subject for further exploration.


\appendix
\section{Scaling form of $f(t;\tq)$ in the Gaussian theory with Model A dynamics}
\label{App:dplusonef}
To determine the generating functional in one higher dimension, we follow the three steps prescribed in \cite{Martin1973,Hochberg1999}:
\begin{itemize}
\item Define $\phi_\zeta$ as the solution of the equation of motion in Eq.~\eqref{Eq:ModelA} for a given noise history $\zeta$ : $\mathcal{J}\phi_\zeta(\vct{k},t) = \zeta(\vct{k},t)$. $\mathcal{J}$ is a linear operator in the Gaussian theory.
\item Rewrite $Z$ as  $\int d\zeta P(\zeta) \exp(J\phi_\zeta)$. The noise distribution, $P(\zeta)$, is Gaussian.
\item Recognize that the probability distribution for $\phi_\zeta$ is related to the noise distribution as $P_{\phi} (\phi_\zeta) = P(\mathcal{J}\phi_\zeta) \mathrm{det}(\mathcal{J})$.
\end{itemize}
The generating functional of correlation functions of $\phi$ so obtained is
\begin{align}
Z[J,\tq]= \int \mathcal{D}\phi \,\textrm{det}(\mathcal{J}) \,e^{\int d^3 k dt \,(-2|\mathcal{J}\phi(\vct{k},t)|^2 + J(\vct{k},t)\phi(-\vct{k},t))}
\end{align}
where
\begin{align}\label{Eq:DDef}
\mathcal{J} =\frac{\partial}{\partial t} + k^2 + r_0(t;\tq) \,.
\end{align}
The functional integral is Gaussian, and the free energy density is expressed in terms of the structure factor in Eq.~\eqref{Eq:GaussianG} as
\begin{equation*}
f(t;\tq) = \int \frac{d^3k}{(2\pi)^3} \log\left[ G_{\sss \phi\phi}^{-1}(k,t;\tq) \right] \,.
\end{equation*}
The challenge in identifying $f_{\rm na}$, even in the time-independent setting, lies in subtracting cut-off dependent terms from $f$ that are analytic in $\delta$. Here, it involves subtracting the cut-off dependent equilibrium contribution at the critical point, $f(0,\infty)$, and two terms that are linear and logarithmic in the cut-off. On taking the KZ scaling limit of the terms remaining, we confirm the scaling form in Eq.~\eqref{Eq:fscalingform}:
\begin{align*}
f_{\rm na}(t;\tq) &\sim \frac{1}{\lk^d} \mathcal{F}(\scale{t}) \\
\mathcal{F}(\scale{t})&= \frac{1}{6\pi^2} \int_0^\infty  d\scale{k} \left(\scale{k}^2\left(2 + \frac{\scale{k}}{\mathcal{G}_{\sss \phi\phi}} \frac{d\mathcal{G}_{\sss \phi\phi}}{dk}\right) + 2  \left(-\scale{t}\right)^a \right) \,.
\end{align*}

\begin{acknowledgements}
\noindent We would like to thank Michael Kolodrubetz for bringing Ref.~\cite{Deng-S.:2008aa} to our attention and a referee for pointing us to Ref.~\cite{De-Grandi:2011lq}. We would also like to thank Tanmay Vachaspati and Nigel Goldenfeld for useful discussions. The work of A.E.\ , A.C.\ and S.L.S.\ was supported in part by the ISF and by NSF grants DMR-1006608 and PHY-1005429.  The work of S.S.G.\ is supported in part by the Department of Energy under Grant No.~DE-FG02-91ER40671.  A.E. is supported by the Adams Fellowship Program of the Israel Academy of Sciences and Humanities.
\end{acknowledgements}

\raggedright

\bibliography{KZBibliography}

\begin{thebibliography}{10}%
\makeatletter
\providecommand \@ifxundefined [1]{%
 \ifx #1\undefined \expandafter \@firstoftwo
 \else \expandafter \@secondoftwo
\fi
}%
\providecommand \@ifnum [1]{%
 \ifnum #1\expandafter \@firstoftwo
 \else \expandafter \@secondoftwo
\fi
}%
\providecommand \enquote [1]{``#1''}%
\providecommand \bibnamefont  [1]{#1}%
\providecommand \bibfnamefont [1]{#1}%
\providecommand \citenamefont [1]{#1}%
\providecommand\href[0]{\@sanitize\@href}%
\providecommand\@href[1]{\endgroup\@@startlink{#1}\endgroup\@@href}%
\providecommand\@@href[1]{#1\@@endlink}%
\providecommand \@sanitize [0]{\begingroup\catcode`\&12\catcode`\#12\relax}%
\@ifxundefined \pdfoutput {\@firstoftwo}{%
 \@ifnum{\z@=\pdfoutput}{\@firstoftwo}{\@secondoftwo}%
}{%
 \providecommand\@@startlink[1]{\leavevmode\special{html:<a href="#1">}}%
 \providecommand\@@endlink[0]{\special{html:</a>}}%
}{%
 \providecommand\@@startlink[1]{%
  \leavevmode
  \pdfstartlink
   attr{/Border[0 0 1 ]/H/I/C[0 1 1]}%
   user{/Subtype/Link/A<</Type/Action/S/URI/URI(#1)>>}%
  \relax
 }%
 \providecommand\@@endlink[0]{\pdfendlink}%
}%
\providecommand \url  [0]{\begingroup\@sanitize \@url }%
\providecommand \@url [1]{\endgroup\@href {#1}{\urlprefix}}%
\providecommand \urlprefix [0]{URL }%
\providecommand \Eprint[0]{\href }%
\@ifxundefined \urlstyle {%
  \providecommand \doi [1]{doi:\discretionary{}{}{}#1}%
}{%
  \providecommand \doi [0]{doi:\discretionary{}{}{}\begingroup
  \urlstyle{rm}\Url }%
}%
\providecommand \doibase [0]{http://dx.doi.org/}%
\providecommand \Doi[1]{\href{\doibase#1}}%
\providecommand \bibAnnote [3]{%
  \BibitemShut{#1}%
  \begin{quotation}\noindent
    \textsc{Key:}\ #2\\\textsc{Annotation:}\ #3%
  \end{quotation}%
}%
\providecommand \bibAnnoteFile [2]{%
  \IfFileExists{#2}{\bibAnnote {#1} {#2} {\input{#2}}}{}%
}%
\providecommand \typeout [0]{\immediate \write \m@ne }%
\providecommand \selectlanguage [0]{\@gobble}%
\providecommand \bibinfo [0]{\@secondoftwo}%
\providecommand \bibfield [0]{\@secondoftwo}%
\providecommand \translation [1]{[#1]}%
\providecommand \BibitemOpen[0]{}%
\providecommand \bibitemStop [0]{}%
\providecommand \bibitemNoStop [0]{.\EOS\space}%
\providecommand \EOS [0]{\spacefactor3000\relax}%
\providecommand \BibitemShut [1]{\csname bibitem#1\endcsname}%
\bibitem{Goldenfeld:1992aa}%
  \BibitemOpen
  \bibfield{author}{%
  \bibinfo {author} {\bibfnamefont{N.}~\bibnamefont{Goldenfeld}},\ }%
  \emph{\bibinfo {title} {{Lectures on Phase Transitions and the
  Renormalization Group}}}\ (\bibinfo {publisher} {Addison-Wesley},\ \bibinfo
  {address} {Mass.},\ \bibinfo {year} {1992})%
  \bibAnnoteFile{NoStop}{Goldenfeld:1992aa}%
\bibitem{Sondhi:1997aa}%
  \BibitemOpen
  \bibfield{author}{%
  \bibinfo {author} {\bibfnamefont{S.~L.}\ \bibnamefont{Sondhi}}, \bibinfo
  {author} {\bibfnamefont{S.~M.}\ \bibnamefont{Girvin}}, \bibinfo {author}
  {\bibfnamefont{J.~P.}\ \bibnamefont{Carini}},\ and\ \bibinfo {author}
  {\bibfnamefont{D.}~\bibnamefont{Shahar}},\ }%
  \bibfield{journal}{%
  \bibinfo {journal} {Rev. Mod. Phys.}\ }%
  \textbf{\bibinfo {volume} {69}},\ \bibinfo {pages} {315} (\bibinfo {month}
  {Jan}\ \bibinfo {year} {1997})%
  \bibAnnoteFile{NoStop}{Sondhi:1997aa}%
\bibitem{Sachdev:1999aa}%
  \BibitemOpen
  \bibfield{author}{%
  \bibinfo {author} {\bibfnamefont{S.}~\bibnamefont{{Sachdev}}},\ }%
  \emph{\bibinfo {title} {Quantum Phase Transitions}}\ (\bibinfo {publisher}
  {Cambridge University Press, Cambridge},\ \bibinfo {year} {1999})%
  \bibAnnoteFile{NoStop}{Sachdev:1999aa}%
\bibitem{Di-Francesco:1999aa}%
  \BibitemOpen
  \bibfield{author}{%
  \bibinfo {author} {\bibfnamefont{P.}~\bibnamefont{Di~Francesco}}, \bibinfo
  {author} {\bibfnamefont{P.}~\bibnamefont{Mathieu}},\ and\ \bibinfo {author}
  {\bibfnamefont{D.}~\bibnamefont{Senechal}},\ }%
  \emph{\bibinfo {title} {Conformal Field Theory}}\ (\bibinfo {publisher}
  {Springer},\ \bibinfo {year} {1999})\ ISBN \bibinfo {isbn} {038794785X}%
  \bibAnnoteFile{NoStop}{Di-Francesco:1999aa}%
\bibitem{Aharony:2000aa}%
  \BibitemOpen
  \bibfield{author}{%
  \bibinfo {author} {\bibfnamefont{O.}~\bibnamefont{Aharony}}, \bibinfo
  {author} {\bibfnamefont{S.~S.}\ \bibnamefont{Gubser}}, \bibinfo {author}
  {\bibfnamefont{J.}~\bibnamefont{Maldacena}}, \bibinfo {author}
  {\bibfnamefont{H.}~\bibnamefont{Ooguri}},\ and\ \bibinfo {author}
  {\bibfnamefont{Y.}~\bibnamefont{Oz}},\ }%
  \bibfield{journal}{%
  \bibinfo {journal} {Physics Reports}\ }%
  \textbf{\bibinfo {volume} {323}},\ \bibinfo {pages} {183 } (\bibinfo {year}
  {2000}),\ ISSN \bibinfo {issn} {0370-1573}%
  \bibAnnoteFile{NoStop}{Aharony:2000aa}%
\bibitem{Kibble1976}%
  \BibitemOpen
  \bibfield{author}{%
  \bibinfo {author} {\bibfnamefont{T.~W.~B.}\ \bibnamefont{Kibble}},\ }%
  \bibfield{journal}{%
  \bibinfo {journal} {Journal of Physics A: Mathematical and General}\ }%
  \textbf{\bibinfo {volume} {9}},\ \bibinfo {pages} {1387} (\bibinfo {year}
  {1976})%
  \bibAnnoteFile{NoStop}{Kibble1976}%
\bibitem{Zurek1985}%
  \BibitemOpen
  \bibfield{author}{%
  \bibinfo {author} {\bibfnamefont{W.~H.}\ \bibnamefont{Zurek}},\ }%
  \bibfield{journal}{%
  \bibinfo {journal} {Nature}\ }%
  \textbf{\bibinfo {volume} {317}},\ \bibinfo {pages} {505} (\bibinfo {month}
  {Oct.}\ \bibinfo {year} {1985})%
  \bibAnnoteFile{NoStop}{Zurek1985}%
\bibitem{Zurek1996}%
  \BibitemOpen
  \bibfield{author}{%
  \bibinfo {author} {\bibfnamefont{W.~H.}\ \bibnamefont{Zurek}},\ }%
  \bibfield{journal}{%
  \bibinfo {journal} {Physics Reports}\ }%
  \textbf{\bibinfo {volume} {276}},\ \bibinfo {pages} {177} (\bibinfo {month}
  {Nov.}\ \bibinfo {year} {1996}),\ ISSN \bibinfo {issn} {0370-1573}%
  \bibAnnoteFile{NoStop}{Zurek1996}%
\bibitem{Chen:2011aa}%
  \BibitemOpen
  \bibfield{author}{%
  \bibinfo {author} {\bibfnamefont{D.}~\bibnamefont{Chen}}, \bibinfo {author}
  {\bibfnamefont{M.}~\bibnamefont{White}}, \bibinfo {author}
  {\bibfnamefont{C.}~\bibnamefont{Borries}},\ and\ \bibinfo {author}
  {\bibfnamefont{B.}~\bibnamefont{DeMarco}},\ }%
  \bibfield{journal}{%
  \Doi{10.1103/PhysRevLett.106.235304}{\bibinfo {journal} {Phys. Rev. Lett.}}\
  }%
  \textbf{\bibinfo {volume} {106}},\ \bibinfo {pages} {235304} (\bibinfo
  {month} {Jun}\ \bibinfo {year} {2011})%
  \bibAnnoteFile{NoStop}{Chen:2011aa}%
\bibitem{Hendry:1994aa}%
  \BibitemOpen
  \bibfield{author}{%
  \bibinfo {author} {\bibfnamefont{P.~C.}\ \bibnamefont{Hendry}}, \bibinfo
  {author} {\bibfnamefont{N.~S.}\ \bibnamefont{Lawson}}, \bibinfo {author}
  {\bibfnamefont{R.~A.~M.}\ \bibnamefont{Lee}}, \bibinfo {author}
  {\bibfnamefont{P.~V.~E.}\ \bibnamefont{McClintock}},\ and\ \bibinfo {author}
  {\bibfnamefont{C.~D.~H.}\ \bibnamefont{Williams}},\ }%
  \bibfield{journal}{%
  \bibinfo {journal} {Nature}\ }%
  \textbf{\bibinfo {volume} {368}},\ \bibinfo {pages} {315} (\bibinfo {month}
  {03}\ \bibinfo {year} {1994})%
  \bibAnnote{NoStop}{Hendry:1994aa}{10.1038/368315a0.}%
\bibitem{Ruutu:1996aa}%
  \BibitemOpen
  \bibfield{author}{%
  \bibinfo {author} {\bibfnamefont{V.~M.~H.}\ \bibnamefont{Ruutu}}, \bibinfo
  {author} {\bibfnamefont{V.~B.}\ \bibnamefont{Eltsov}}, \bibinfo {author}
  {\bibfnamefont{A.~J.}\ \bibnamefont{Gill}}, \bibinfo {author}
  {\bibfnamefont{T.~W.~B.}\ \bibnamefont{Kibble}}, \bibinfo {author}
  {\bibfnamefont{M.}~\bibnamefont{Krusius}}, \bibinfo {author}
  {\bibfnamefont{Y.~G.}\ \bibnamefont{Makhlin}}, \bibinfo {author}
  {\bibfnamefont{B.}~\bibnamefont{Placais}}, \bibinfo {author}
  {\bibfnamefont{G.~E.}\ \bibnamefont{Volovik}},\ and\ \bibinfo {author}
  {\bibfnamefont{W.}~\bibnamefont{Xu}},\ }%
  \bibfield{journal}{%
  \bibinfo {journal} {Nature}\ }%
  \textbf{\bibinfo {volume} {382}},\ \bibinfo {pages} {334} (\bibinfo {month}
  {07}\ \bibinfo {year} {1996})%
  \bibAnnote{NoStop}{Ruutu:1996aa}{10.1038/382334a0.}%
\bibitem{Bauerle:1996aa}%
  \BibitemOpen
  \bibfield{author}{%
  \bibinfo {author} {\bibfnamefont{C.}~\bibnamefont{Bauerle}}, \bibinfo
  {author} {\bibfnamefont{Y.~M.}\ \bibnamefont{Bunkov}}, \bibinfo {author}
  {\bibfnamefont{S.~N.}\ \bibnamefont{Fisher}}, \bibinfo {author}
  {\bibfnamefont{H.}~\bibnamefont{Godfrin}},\ and\ \bibinfo {author}
  {\bibfnamefont{G.~R.}\ \bibnamefont{Pickett}},\ }%
  \bibfield{journal}{%
  \bibinfo {journal} {Nature}\ }%
  \textbf{\bibinfo {volume} {382}},\ \bibinfo {pages} {332} (\bibinfo {month}
  {07}\ \bibinfo {year} {1996})%
  \bibAnnote{NoStop}{Bauerle:1996aa}{10.1038/382332a0.}%
\bibitem{Maniv2003}%
  \BibitemOpen
  \bibfield{author}{%
  \bibinfo {author} {\bibfnamefont{A.}~\bibnamefont{Maniv}}, \bibinfo {author}
  {\bibfnamefont{E.}~\bibnamefont{Polturak}},\ and\ \bibinfo {author}
  {\bibfnamefont{G.}~\bibnamefont{Koren}},\ }%
  \bibfield{journal}{%
  \bibinfo {journal} {Phys. Rev. Lett.}\ }%
  \textbf{\bibinfo {volume} {91}},\ \bibinfo {pages} {197001} (\bibinfo {month}
  {Nov.}\ \bibinfo {year} {2003})%
  \bibAnnoteFile{NoStop}{Maniv2003}%
\bibitem{Monaco:2006ly}%
  \BibitemOpen
  \bibfield{author}{%
  \bibinfo {author} {\bibfnamefont{R.}~\bibnamefont{Monaco}}, \bibinfo {author}
  {\bibfnamefont{J.}~\bibnamefont{Mygind}}, \bibinfo {author}
  {\bibfnamefont{M.}~\bibnamefont{Aaroe}}, \bibinfo {author}
  {\bibfnamefont{R.~J.}\ \bibnamefont{Rivers}},\ and\ \bibinfo {author}
  {\bibfnamefont{V.~P.}\ \bibnamefont{Koshelets}},\ }%
  \bibfield{journal}{%
  \Doi{10.1103/PhysRevLett.96.180604}{\bibinfo {journal} {Phys. Rev. Lett.}}\
  }%
  \textbf{\bibinfo {volume} {96}},\ \bibinfo {pages} {180604} (\bibinfo {month}
  {May}\ \bibinfo {year} {2006}),\
  \url{http://link.aps.org/doi/10.1103/PhysRevLett.96.180604}%
  \bibAnnoteFile{NoStop}{Monaco:2006ly}%
\bibitem{Ducci:1999aa}%
  \BibitemOpen
  \bibfield{author}{%
  \bibinfo {author} {\bibfnamefont{S.}~\bibnamefont{Ducci}}, \bibinfo {author}
  {\bibfnamefont{P.~L.}\ \bibnamefont{Ramazza}}, \bibinfo {author}
  {\bibfnamefont{W.}~\bibnamefont{Gonz\'alez-Vi\~nas}},\ and\ \bibinfo {author}
  {\bibfnamefont{F.~T.}\ \bibnamefont{Arecchi}},\ }%
  \bibfield{journal}{%
  \bibinfo {journal} {Phys. Rev. Lett.}\ }%
  \textbf{\bibinfo {volume} {83}},\ \bibinfo {pages} {5210} (\bibinfo {month}
  {Dec}\ \bibinfo {year} {1999})%
  \bibAnnoteFile{NoStop}{Ducci:1999aa}%
\bibitem{Casado:2006aa}%
  \BibitemOpen
  \bibfield{author}{%
  \bibinfo {author} {\bibfnamefont{S.}~\bibnamefont{Casado}}, \bibinfo {author}
  {\bibfnamefont{W.}~\bibnamefont{Gonz\'alez-Vi\~nas}},\ and\ \bibinfo {author}
  {\bibfnamefont{H.}~\bibnamefont{Mancini}},\ }%
  \bibfield{journal}{%
  \Doi{10.1103/PhysRevE.74.047101}{\bibinfo {journal} {Phys. Rev. E}}\ }%
  \textbf{\bibinfo {volume} {74}},\ \bibinfo {pages} {047101} (\bibinfo {month}
  {Oct}\ \bibinfo {year} {2006})%
  \bibAnnoteFile{NoStop}{Casado:2006aa}%
\bibitem{Dziarmaga:2005aa}%
  \BibitemOpen
  \bibfield{author}{%
  \bibinfo {author} {\bibfnamefont{J.}~\bibnamefont{Dziarmaga}},\ }%
  \bibfield{journal}{%
  \bibinfo {journal} {Phys. Rev. Lett.}\ }%
  \textbf{\bibinfo {volume} {95}},\ \bibinfo {pages} {245701} (\bibinfo {month}
  {Dec}\ \bibinfo {year} {2005})%
  \bibAnnoteFile{NoStop}{Dziarmaga:2005aa}%
\bibitem{Dziarmaga:2010aa}%
  \BibitemOpen
  \bibfield{author}{%
  \bibinfo {author} {\bibfnamefont{J.}~\bibnamefont{Dziarmaga}},\ }%
  \bibfield{journal}{%
  \bibinfo {journal} {Advances in Physics}\ }%
  \textbf{\bibinfo {volume} {59}},\ \bibinfo {pages} {1063} (\bibinfo {year}
  {2010})%
  \bibAnnoteFile{NoStop}{Dziarmaga:2010aa}%
\bibitem{Polkovnikov:2005aa}%
  \BibitemOpen
  \bibfield{author}{%
  \bibinfo {author} {\bibfnamefont{A.}~\bibnamefont{Polkovnikov}},\ }%
  \bibfield{journal}{%
  \bibinfo {journal} {Phys. Rev. B}\ }%
  \textbf{\bibinfo {volume} {72}} (\bibinfo {year} {2005})%
  \bibAnnoteFile{NoStop}{Polkovnikov:2005aa}%
\bibitem{Zurek:2005aa}%
  \BibitemOpen
  \bibfield{author}{%
  \bibinfo {author} {\bibfnamefont{W.~H.}\ \bibnamefont{Zurek}}, \bibinfo
  {author} {\bibfnamefont{U.}~\bibnamefont{Dorner}},\ and\ \bibinfo {author}
  {\bibfnamefont{P.}~\bibnamefont{Zoller}},\ }%
  \bibfield{journal}{%
  \bibinfo {journal} {Phys. Rev. Lett.}\ }%
  \textbf{\bibinfo {volume} {95}},\ \bibinfo {pages} {105701} (\bibinfo {month}
  {Sep}\ \bibinfo {year} {2005})%
  \bibAnnoteFile{NoStop}{Zurek:2005aa}%
\bibitem{Divakaran:2009fk}%
  \BibitemOpen
  \bibfield{author}{%
  \bibinfo {author} {\bibfnamefont{U.}~\bibnamefont{Divakaran}}, \bibinfo
  {author} {\bibfnamefont{V.}~\bibnamefont{Mukherjee}}, \bibinfo {author}
  {\bibfnamefont{A.}~\bibnamefont{Dutta}},\ and\ \bibinfo {author}
  {\bibfnamefont{D.}~\bibnamefont{Sen}},\ }%
  \bibfield{journal}{%
  \bibinfo {journal} {Journal of Statistical Mechanics: Theory and Experiment}\
  }%
  \textbf{\bibinfo {volume} {2009}},\ \bibinfo {pages} {P02007} (\bibinfo
  {year} {2009}),\ \url{http://stacks.iop.org/1742-5468/2009/i=02/a=P02007}%
  \bibAnnoteFile{NoStop}{Divakaran:2009fk}%
\bibitem{Divakaran:2008kx}%
  \BibitemOpen
  \bibfield{author}{%
  \bibinfo {author} {\bibfnamefont{U.}~\bibnamefont{Divakaran}}, \bibinfo
  {author} {\bibfnamefont{A.}~\bibnamefont{Dutta}},\ and\ \bibinfo {author}
  {\bibfnamefont{D.}~\bibnamefont{Sen}},\ }%
  \bibfield{journal}{%
  \Doi{10.1103/PhysRevB.78.144301}{\bibinfo {journal} {Phys. Rev. B}}\ }%
  \textbf{\bibinfo {volume} {78}},\ \bibinfo {pages} {144301} (\bibinfo {month}
  {Oct}\ \bibinfo {year} {2008}),\
  \url{http://link.aps.org/doi/10.1103/PhysRevB.78.144301}%
  \bibAnnoteFile{NoStop}{Divakaran:2008kx}%
\bibitem{De-Grandi:2010aa}%
  \BibitemOpen
  \bibfield{author}{%
  \bibinfo {author} {\bibfnamefont{C.}~\bibnamefont{De~Grandi}}, \bibinfo
  {author} {\bibfnamefont{V.}~\bibnamefont{Gritsev}},\ and\ \bibinfo {author}
  {\bibfnamefont{A.}~\bibnamefont{Polkovnikov}},\ }%
  \bibfield{journal}{%
  \bibinfo {journal} {Phys. Rev. B}\ }%
  \textbf{\bibinfo {volume} {81}},\ \bibinfo {pages} {012303} (\bibinfo {month}
  {Jan}\ \bibinfo {year} {2010})%
  \bibAnnoteFile{NoStop}{De-Grandi:2010aa}%
\bibitem{Polkovnikov:2010aa}%
  \BibitemOpen
  \bibfield{author}{%
  \bibinfo {author} {\bibfnamefont{A.}~\bibnamefont{Polkovnikov}}, \bibinfo
  {author} {\bibfnamefont{K.}~\bibnamefont{Sengupta}}, \bibinfo {author}
  {\bibfnamefont{A.}~\bibnamefont{Silva}},\ and\ \bibinfo {author}
  {\bibfnamefont{M.}~\bibnamefont{Vengalattore}},\ }%
  \bibfield{journal}{%
  \Doi{10.1103/RevModPhys.83.863}{\bibinfo {journal} {Rev. Mod. Phys.}}\ }%
  \textbf{\bibinfo {volume} {83}},\ \bibinfo {pages} {863} (\bibinfo {month}
  {Aug}\ \bibinfo {year} {2011}),\
  \url{http://link.aps.org/doi/10.1103/RevModPhys.83.863}%
  \bibAnnoteFile{NoStop}{Polkovnikov:2010aa}%
\bibitem{Deng-S.:2008aa}%
  \BibitemOpen
  \bibfield{author}{%
  \bibinfo {author} {\bibnamefont{{Deng, S.}}}, \bibinfo {author}
  {\bibnamefont{{Ortiz, G.}}},\ and\ \bibinfo {author} {\bibnamefont{{Viola,
  L.}}},\ }%
  \bibfield{journal}{%
  \bibinfo {journal} {EPL}\ }%
  \textbf{\bibinfo {volume} {84}},\ \bibinfo {pages} {67008} (\bibinfo {year}
  {2008})%
  \bibAnnoteFile{NoStop}{Deng-S.:2008aa}%
\bibitem{Biroli:2010aa}%
  \BibitemOpen
  \bibfield{author}{%
  \bibinfo {author} {\bibfnamefont{G.}~\bibnamefont{Biroli}}, \bibinfo {author}
  {\bibfnamefont{L.~F.}\ \bibnamefont{Cugliandolo}},\ and\ \bibinfo {author}
  {\bibfnamefont{A.}~\bibnamefont{Sicilia}},\ }%
  \bibfield{journal}{%
  \bibinfo {journal} {Phys. Rev. E}\ }%
  \textbf{\bibinfo {volume} {81}},\ \bibinfo {pages} {050101} (\bibinfo {month}
  {May}\ \bibinfo {year} {2010})%
  \bibAnnoteFile{NoStop}{Biroli:2010aa}%
\bibitem{De-Grandi:2011lq}%
  \BibitemOpen
  \bibfield{author}{%
  \bibinfo {author} {\bibfnamefont{C.}~\bibnamefont{De~Grandi}}, \bibinfo
  {author} {\bibfnamefont{A.}~\bibnamefont{Polkovnikov}},\ and\ \bibinfo
  {author} {\bibfnamefont{A.~W.}\ \bibnamefont{Sandvik}},\ }%
  \bibfield{journal}{%
  \Doi{10.1103/PhysRevB.84.224303}{\bibinfo {journal} {Phys. Rev. B}}\ }%
  \textbf{\bibinfo {volume} {84}},\ \bibinfo {pages} {224303} (\bibinfo {month}
  {Dec}\ \bibinfo {year} {2011}),\
  \url{http://link.aps.org/doi/10.1103/PhysRevB.84.224303}%
  \bibAnnoteFile{NoStop}{De-Grandi:2011lq}%
\bibitem{Calabrese:2005ab}%
  \BibitemOpen
  \bibfield{author}{%
  \bibinfo {author} {\bibfnamefont{P.}~\bibnamefont{Calabrese}}\ and\ \bibinfo
  {author} {\bibfnamefont{A.}~\bibnamefont{Gambassi}},\ }%
  \bibfield{journal}{%
  \bibinfo {journal} {Journal of Physics A: Mathematical and General}\ }%
  \textbf{\bibinfo {volume} {38}},\ \bibinfo {pages} {R133} (\bibinfo {year}
  {2005})%
  \bibAnnoteFile{NoStop}{Calabrese:2005ab}%
\bibitem{Sen:2008aa}%
  \BibitemOpen
  \bibfield{author}{%
  \bibinfo {author} {\bibfnamefont{D.}~\bibnamefont{Sen}}, \bibinfo {author}
  {\bibfnamefont{K.}~\bibnamefont{Sengupta}},\ and\ \bibinfo {author}
  {\bibfnamefont{S.}~\bibnamefont{Mondal}},\ }%
  \bibfield{journal}{%
  \bibinfo {journal} {Phys. Rev. Lett.}\ }%
  \textbf{\bibinfo {volume} {101}},\ \bibinfo {pages} {016806} (\bibinfo
  {month} {Jul}\ \bibinfo {year} {2008})%
  \bibAnnoteFile{NoStop}{Sen:2008aa}%
\bibitem{Kolodrubetz:2011fj}%
  \BibitemOpen
  \bibfield{author}{%
  \bibinfo {author} {\bibfnamefont{M.}~\bibnamefont{{Kolodrubetz}}}, \bibinfo
  {author} {\bibfnamefont{B.~K.}\ \bibnamefont{{Clark}}},\ and\ \bibinfo
  {author} {\bibfnamefont{D.~A.}\ \bibnamefont{{Huse}}}}%
   (\bibinfo {month} {Dec.}\ \bibinfo {year} {2011}),\
  \Eprint{http://arxiv.org/abs/1112.6422}{arXiv:1112.6422}%
  \bibAnnoteFile{NoStop}{Kolodrubetz:2011fj}%
\bibitem{Note1}%
  \BibitemOpen
  \bibinfo {note} {Anticipating our classification, this statement is strictly
  true only for the cis- and end-critical protocols. The dangerously irrelevant
  interaction needs to be included to properly study the trans-critical
  protocols.}%
  \bibAnnoteFile{Stop}{Note1}%
\bibitem{Chandran:2012aa}%
  \BibitemOpen
  \bibfield{author}{%
  \bibinfo {author} {\bibfnamefont{A.}~\bibnamefont{Chandran}}, \bibinfo
  {author} {\bibfnamefont{A.}~\bibnamefont{Erez}}, \bibinfo {author}
  {\bibfnamefont{S.}~\bibnamefont{Gubser}},\ and\ \bibinfo {author}
  {\bibfnamefont{S.}~\bibnamefont{Sondhi}},\ }%
  \bibinfo {note} {in preparation}%
  \bibAnnoteFile{NoStop}{Chandran:2012aa}%
\bibitem{Note2}%
  \BibitemOpen
  \bibinfo {note} {Near a quantum phase transition, this implies a vanishing
  many-body gap.}%
  \bibAnnoteFile{Stop}{Note2}%
\bibitem{Note3}%
  \BibitemOpen
  \bibinfo {note} {With weak disorder, the KZ scaling forms apply to disorder
  averaged connected correlation functions. When the disorder distributions are
  broad, the typical moments satisfy the proposed scaling.}%
  \bibAnnoteFile{Stop}{Note3}%
\bibitem{Note4}%
  \BibitemOpen
  \bibinfo {note} {For a recent dissent, see Olejarz et al.\ in Phys. Rev. E
  \protect \textbf {83}, 05144 (2011) who have noted that the 3d Ising Model
  does not coarsen at zero temperature. More generally, the KZ scaling forms
  asymptote to the long-time behavior in the sudden quench.}%
  \bibAnnoteFile{Stop}{Note4}%
\bibitem{Bray1994}%
  \BibitemOpen
  \bibfield{author}{%
  \bibinfo {author} {\bibfnamefont{A.~J.}\ \bibnamefont{Bray}},\ }%
  \bibfield{journal}{%
  \bibinfo {journal} {Advances in Physics}\ }%
  \textbf{\bibinfo {volume} {43}},\ \bibinfo {pages} {357 } (\bibinfo {year}
  {1994})%
  \bibAnnoteFile{NoStop}{Bray1994}%
\bibitem{Chuang:1991aa}%
  \BibitemOpen
  \bibfield{author}{%
  \bibinfo {author} {\bibfnamefont{I.}~\bibnamefont{Chuang}}, \bibinfo {author}
  {\bibfnamefont{R.}~\bibnamefont{Durrer}}, \bibinfo {author}
  {\bibfnamefont{N.}~\bibnamefont{Turok}},\ and\ \bibinfo {author}
  {\bibfnamefont{B.}~\bibnamefont{Yurke}},\ }%
  \bibfield{journal}{%
  \bibinfo {journal} {Science}\ }%
  \textbf{\bibinfo {volume} {251}},\ \bibinfo {pages} {1336} (\bibinfo {year}
  {1991})%
  \bibAnnoteFile{NoStop}{Chuang:1991aa}%
\bibitem{Bowick:1994aa}%
  \BibitemOpen
  \bibfield{author}{%
  \bibinfo {author} {\bibfnamefont{M.~J.}\ \bibnamefont{Bowick}}, \bibinfo
  {author} {\bibfnamefont{L.}~\bibnamefont{Chandar}}, \bibinfo {author}
  {\bibfnamefont{E.~A.}\ \bibnamefont{Schiff}},\ and\ \bibinfo {author}
  {\bibfnamefont{A.~M.}\ \bibnamefont{Srivastava}},\ }%
  \bibfield{journal}{%
  \bibinfo {journal} {Science}\ }%
  \textbf{\bibinfo {volume} {263}},\ \bibinfo {pages} {943} (\bibinfo {month}
  {02}\ \bibinfo {year} {1994})%
  \bibAnnote{NoStop}{Bowick:1994aa}{The production of strings (disclination
  lines and loops) has been observed by means of the Kibble mechanism of domain
  (bubble) formation in the isotropic-nematic phase transition of the uniaxial
  nematic liquid crystal 4-cyano-4′-n-pentylbiphenyl. The number of strings
  formed per bubble is about 0.6. This value is in reasonable agreement with a
  numerical simulation of the experiment in which the Kibble mechanism is used
  for the order parameter space of a uniaxial nematic liquid crystal.}%
\bibitem{Halperin:1981aa}%
  \BibitemOpen
  \bibfield{author}{%
  \bibinfo {author} {\bibfnamefont{B.~I.}\ \bibnamefont{Halperin}},\ }%
  \emph{\bibinfo {title} {Physics of Defects}},\ Proceedings of the Les Houches
  Summer Institute\ (\bibinfo {publisher} {North Holland, Amsterdam},\ \bibinfo
  {year} {1981})%
  \bibAnnoteFile{NoStop}{Halperin:1981aa}%
\bibitem{Liu:1992fk}%
  \BibitemOpen
  \bibfield{author}{%
  \bibinfo {author} {\bibfnamefont{F.}~\bibnamefont{Liu}}\ and\ \bibinfo
  {author} {\bibfnamefont{G.~F.}\ \bibnamefont{Mazenko}},\ }%
  \bibfield{journal}{%
  \bibinfo {journal} {Phys. Rev. B}\ }%
  \textbf{\bibinfo {volume} {46}},\ \bibinfo {pages} {5963} (\bibinfo {month}
  {Sep}\ \bibinfo {year} {1992})%
  \bibAnnoteFile{NoStop}{Liu:1992fk}%
\bibitem{Hohenberg:1977aa}%
  \BibitemOpen
  \bibfield{author}{%
  \bibinfo {author} {\bibfnamefont{P.~C.}\ \bibnamefont{Hohenberg}}\ and\
  \bibinfo {author} {\bibfnamefont{B.~I.}\ \bibnamefont{Halperin}},\ }%
  \bibfield{journal}{%
  \bibinfo {journal} {Rev. Mod. Phys.}\ }%
  \textbf{\bibinfo {volume} {49}},\ \bibinfo {pages} {435} (\bibinfo {month}
  {Jul}\ \bibinfo {year} {1977})%
  \bibAnnoteFile{NoStop}{Hohenberg:1977aa}%
\bibitem{Hochberg1999}%
  \BibitemOpen
  \bibfield{author}{%
  \bibinfo {author} {\bibfnamefont{D.}~\bibnamefont{Hochberg}}, \bibinfo
  {author} {\bibfnamefont{C.}~\bibnamefont{Molina-Par\'\i{}s}}, \bibinfo
  {author} {\bibfnamefont{J.}~\bibnamefont{P\'erez-Mercader}},\ and\ \bibinfo
  {author} {\bibfnamefont{M.}~\bibnamefont{Visser}},\ }%
  \bibfield{journal}{%
  \bibinfo {journal} {Phys. Rev. E}\ }%
  \textbf{\bibinfo {volume} {60}},\ \bibinfo {pages} {6343} (\bibinfo {month}
  {Dec}\ \bibinfo {year} {1999})%
  \bibAnnoteFile{NoStop}{Hochberg1999}%
\bibitem{Martin1973}%
  \BibitemOpen
  \bibfield{author}{%
  \bibinfo {author} {\bibfnamefont{P.~C.}\ \bibnamefont{Martin}}, \bibinfo
  {author} {\bibfnamefont{E.~D.}\ \bibnamefont{Siggia}},\ and\ \bibinfo
  {author} {\bibfnamefont{H.~A.}\ \bibnamefont{Rose}},\ }%
  \bibfield{journal}{%
  \bibinfo {journal} {Phys. Rev. A}\ }%
  \textbf{\bibinfo {volume} {8}},\ \bibinfo {pages} {423} (\bibinfo {month}
  {Jul}\ \bibinfo {year} {1973})%
  \bibAnnoteFile{NoStop}{Martin1973}%
\bibitem{Janssen:1989kx}%
  \BibitemOpen
  \bibfield{author}{%
  \bibinfo {author} {\bibfnamefont{H.~K.}\ \bibnamefont{Janssen}}, \bibinfo
  {author} {\bibfnamefont{B.}~\bibnamefont{Schaub}},\ and\ \bibinfo {author}
  {\bibfnamefont{B.}~\bibnamefont{Schmittmann}},\ }%
  \bibfield{journal}{%
  \bibinfo {journal} {Zeitschrift f{\"u}r Physik B Condensed Matter}\ }%
  \textbf{\bibinfo {volume} {73}},\ \bibinfo {pages} {539} (\bibinfo {year}
  {1989}),\ ISSN \bibinfo {issn} {0722-3277},\
  \url{http://dx.doi.org/10.1007/BF01319383}%
  \bibAnnoteFile{NoStop}{Janssen:1989kx}%
\bibitem{Note5}%
  \BibitemOpen
  \bibinfo {note} {See \cite {Polkovnikov:2008ab} for a discussion of why the
  total excitation energy is sensibly called heat. For infinite systems where
  one needs to work with intensive quantities though, a heat density is not a
  useful concept especially when macroscopic subregions exhibit thermal
  equilibration}%
  \bibAnnoteFile{NoStop}{Note5}%
\bibitem{Anatoli:2011aa}%
  \BibitemOpen
  \bibfield{author}{%
  \bibinfo {author} {\bibfnamefont{A.}~\bibnamefont{Polkovnikov}},\ }%
  \bibfield{journal}{%
  \bibinfo {journal} {Annals of Physics}\ }%
  \textbf{\bibinfo {volume} {326}},\ \bibinfo {pages} {486 } (\bibinfo {year}
  {2011}),\ ISSN \bibinfo {issn} {0003-4916}%
  \bibAnnoteFile{NoStop}{Anatoli:2011aa}%
\bibitem{Polkovnikov:2008ab}%
  \BibitemOpen
  \bibfield{author}{%
  \bibinfo {author} {\bibfnamefont{A.}~\bibnamefont{Polkovnikov}},\ }%
  \bibfield{journal}{%
  \bibinfo {journal} {Phys. Rev. Lett.}\ }%
  \textbf{\bibinfo {volume} {101}},\ \bibinfo {pages} {220402} (\bibinfo
  {month} {Nov}\ \bibinfo {year} {2008})%
  \bibAnnoteFile{NoStop}{Polkovnikov:2008ab}%
\bibitem{Note6}%
  \BibitemOpen
  \bibinfo {note} {For an isolated system at finite temperature, the dynamics
  involves starting with a typical state with the thermodynamic limit energy
  density. At high temperatures, a typical state must look ``classical''; this
  suggests that the behavior of the entanglement entropy and issues of
  many-body localization \cite {Basko:2006aa} need further examination.}%
  \bibAnnoteFile{Stop}{Note6}%
\bibitem{Uhlmann:2010aa}%
  \BibitemOpen
  \bibfield{author}{%
  \bibinfo {author} {\bibfnamefont{M.}~\bibnamefont{Uhlmann}}, \bibinfo
  {author} {\bibfnamefont{R.}~\bibnamefont{Sch\"utzhold}},\ and\ \bibinfo
  {author} {\bibfnamefont{U.~R.}\ \bibnamefont{Fischer}},\ }%
  \bibfield{journal}{%
  \bibinfo {journal} {Phys. Rev. D}\ }%
  \textbf{\bibinfo {volume} {81}},\ \bibinfo {pages} {025017} (\bibinfo {month}
  {Jan}\ \bibinfo {year} {2010})%
  \bibAnnoteFile{NoStop}{Uhlmann:2010aa}%
\bibitem{Polkovnikov:2008aa}%
  \BibitemOpen
  \bibfield{author}{%
  \bibinfo {author} {\bibfnamefont{A.}~\bibnamefont{Polkovnikov}}\ and\
  \bibinfo {author} {\bibfnamefont{V.}~\bibnamefont{Gritsev}},\ }%
  \bibfield{journal}{%
  \bibinfo {journal} {Nat Phys}\ }%
  \textbf{\bibinfo {volume} {4}},\ \bibinfo {pages} {477} (\bibinfo {month}
  {06}\ \bibinfo {year} {2008})%
  \bibAnnote{NoStop}{Polkovnikov:2008aa}{10.1038/nphys963.}%
\bibitem{Note7}%
  \BibitemOpen
  \bibinfo {note} {We exclude dangerously irrelevant operators from this
  discussion as they will modify the asymptotic behavior in the KZ scaling
  limit.}%
  \bibAnnoteFile{Stop}{Note7}%
\bibitem{Birrell:1982ix}%
  \BibitemOpen
  \bibfield{author}{%
  \bibinfo {author} {\bibfnamefont{N.~D.}\ \bibnamefont{Birrell}}\ and\
  \bibinfo {author} {\bibfnamefont{P.~C.~W.}\ \bibnamefont{Davies}},\ }%
  \emph{\bibinfo {title} {{Quantum Fields in Curved Space}}}\ (\bibinfo
  {publisher} {Cambridge University Press},\ \bibinfo {year} {1982})%
  \bibAnnoteFile{NoStop}{Birrell:1982ix}%
\bibitem{Kluger1998}%
  \BibitemOpen
  \bibfield{author}{%
  \bibinfo {author} {\bibfnamefont{Y.}~\bibnamefont{Kluger}}, \bibinfo {author}
  {\bibfnamefont{E.}~\bibnamefont{Mottola}},\ and\ \bibinfo {author}
  {\bibfnamefont{J.~M.}\ \bibnamefont{Eisenberg}},\ }%
  \bibfield{journal}{%
  \bibinfo {journal} {Phys. Rev. D}\ }%
  \textbf{\bibinfo {volume} {58}},\ \bibinfo {pages} {125015} (\bibinfo {month}
  {Nov.}\ \bibinfo {year} {1998})%
  \bibAnnoteFile{NoStop}{Kluger1998}%
\bibitem{Dantas:1992aa}%
  \BibitemOpen
  \bibfield{author}{%
  \bibinfo {author} {\bibfnamefont{C.~M.~A.}\ \bibnamefont{Dantas}}, \bibinfo
  {author} {\bibfnamefont{I.~A.}\ \bibnamefont{Pedrosa}},\ and\ \bibinfo
  {author} {\bibfnamefont{B.}~\bibnamefont{Baseia}},\ }%
  \bibfield{journal}{%
  \bibinfo {journal} {Phys. Rev. A}\ }%
  \textbf{\bibinfo {volume} {45}},\ \bibinfo {pages} {1320} (\bibinfo {month}
  {Feb}\ \bibinfo {year} {1992})%
  \bibAnnoteFile{NoStop}{Dantas:1992aa}%
\bibitem{Note8}%
  \BibitemOpen
  \bibinfo {note} {The reader may worry about dropping the second term at the
  zeros of $g_+(kt)$. Fortunately, the correction to $f_k(t)$ due to
  $g_{-}(kt)$ at these points decreases as $t\rightarrow \infty $.}%
  \bibAnnoteFile{Stop}{Note8}%
\bibitem{Strominger:2001pn}%
  \BibitemOpen
  \bibfield{author}{%
  \bibinfo {author} {\bibfnamefont{A.}~\bibnamefont{Strominger}},\ }%
  \bibfield{journal}{%
  \bibinfo {journal} {JHEP}\ }%
  \textbf{\bibinfo {volume} {10}},\ \bibinfo {pages} {034} (\bibinfo {year}
  {2001})%
  \bibAnnoteFile{NoStop}{Strominger:2001pn}%
\bibitem{Witten:2001ua}%
  \BibitemOpen
  \bibfield{author}{%
  \bibinfo {author} {\bibfnamefont{E.}~\bibnamefont{Witten}}}%
   (\bibinfo {year} {2001}),\
  \Eprint{http://arxiv.org/abs/hep-th/0112258}{arXiv:hep-th/0112258}%
  \bibAnnoteFile{NoStop}{Witten:2001ua}%
\bibitem{Breitenlohner:1982jf}%
  \BibitemOpen
  \bibfield{author}{%
  \bibinfo {author} {\bibfnamefont{P.}~\bibnamefont{Breitenlohner}}\ and\
  \bibinfo {author} {\bibfnamefont{D.~Z.}\ \bibnamefont{Freedman}},\ }%
  \bibfield{journal}{%
  \Doi{10.1016/0003-4916(82)90116-6}{\bibinfo {journal} {Ann. Phys.}}\ }%
  \textbf{\bibinfo {volume} {144}},\ \bibinfo {pages} {249} (\bibinfo {year}
  {1982})%
  \bibAnnoteFile{NoStop}{Breitenlohner:1982jf}%
\bibitem{Hindmarsh:2000uq}%
  \BibitemOpen
  \bibfield{author}{%
  \bibinfo {author} {\bibfnamefont{M.}~\bibnamefont{Hindmarsh}}\ and\ \bibinfo
  {author} {\bibfnamefont{A.}~\bibnamefont{Rajantie}},\ }%
  \bibfield{journal}{%
  \Doi{10.1103/PhysRevLett.85.4660}{\bibinfo {journal} {Phys. Rev. Lett.}}\ }%
  \textbf{\bibinfo {volume} {85}},\ \bibinfo {pages} {4660} (\bibinfo {month}
  {Nov}\ \bibinfo {year} {2000})%
  \bibAnnoteFile{NoStop}{Hindmarsh:2000uq}%
\bibitem{Basko:2006aa}%
  \BibitemOpen
  \bibfield{author}{%
  \bibinfo {author} {\bibfnamefont{D.}~\bibnamefont{Basko}}, \bibinfo {author}
  {\bibfnamefont{I.}~\bibnamefont{Aleiner}},\ and\ \bibinfo {author}
  {\bibfnamefont{B.}~\bibnamefont{Altshuler}},\ }%
  \bibfield{journal}{%
  \bibinfo {journal} {Annals of Physics}\ }%
  \textbf{\bibinfo {volume} {321}},\ \bibinfo {pages} {1126 } (\bibinfo {year}
  {2006}),\ ISSN \bibinfo {issn} {0003-4916}%
  \bibAnnoteFile{NoStop}{Basko:2006aa}%
\end{thebibliography}%

\end{document}